\documentclass[%
 reprint,
nofootinbib,
 amsmath,amssymb,
 aps,
twocolumn
]{revtex4-2}
\usepackage{import}
\usepackage[utf8]{inputenc}
\usepackage[capitalize]{cleveref}
\usepackage{graphicx}
\usepackage{dcolumn}
\usepackage{bm}
\usepackage{mathrsfs}

\usepackage{pgfplots}

\pgfplotsset{compat=1.15}
\usepackage{verbatim}
\usepackage{array}
\usepackage{multirow}
\usepackage[export]{adjustbox}
\newcolumntype{L}{>{\centering\arraybackslash}m{1.4cm}}
\newcolumntype{P}[1]{>{\centering\arraybackslash}p{#1}}

\usepackage{etoolbox}
\makeatletter
\patchcmd\linenumberpar{\@LN@parpgbrk}{\penalty\@LN@parpgpen\relax}{}{}
\makeatother

\begin{document}

\renewcommand{\arraystretch}{1.5}
\preprint{APS/123-QED}

\title{Impact of biaxial birefringence in polar ice
at
radio frequencies 
on 
signal polarizations in ultra-high energy neutrino detection  }

\author{Amy Connolly}
\email{connolly@physics.osu.edu}
\affiliation{%
 Department of Physics and Center for Cosmology and AstroParticle Physics (CCAPP), Ohio State University, Columbus, Ohio 43210, USA
}%

\date{\today}

\begin{abstract}
It is known that polar ice is birefringent
and that this
can have implications for in-ice
radio detection of ultra-high energy
neutrinos.
Previous investigations of the effects of birefringence on the propagation of radio-frequency signals in ice have found that it can cause
time delays
between pulses in different polarizations
in in-ice neutrino experiments,
and can have polarization-dependent
effects on power in radar echoes at
oblique angles in polar ice.
I report, for the first time, on
implications for the received power in
different polarizations 
in
high energy neutrino experiments,
where the source of the emitted signal 
is in the ice, a biaxial
treatment at radio wavelengths is
used, and the signals propagate at
oblique angles.
I describe a model for this and 
compare with published results from the SPICE in-ice calibration
 pulser system at South Pole, where unexpectedly
 high cross-polarization power has been reported for
 some geometries.  
The data 
 shows behaviors indicative of the need for a biaxial treatment of 
 birefringence inducing non-trivial rotations
of the signal polarization. 
The behaviors 
include, but are not limited to, a time delay
that would leave 
an imprint in the power spectrum.  I explain why 
this time delay has the potential to serve as both an in-ice neutrino signature and a
measurement of the distance to the interaction.
  While
further work is needed, I 
expect
that proper handling of 
the effects presented here 
will increase the science potential 
of ultra-high energy neutrino experiments, 
and may impact the optimal designs
of next-generation detectors.  
\end{abstract}

\maketitle

\section{\label{sec:introduction}Introduction }

Ultra-high energy neutrinos are a crucial
missing piece in the rapidly expanding
field of multi-messenger astrophysics~\cite{IceCube:2018dnn,IceCube:2018cha,Monitor:2017mdv}.
Alongside the mature and evolving measurements
of cosmic rays and gamma rays up to their
highest detectable energies, 
the past decade has seen the discovery
of a high energy astrophysical neutrino
flux up to O(10)\,PeV~\cite{Aartsen:2013jdh,Aartsen:2015rwa,Aartsen:2013bka}, 
as well as the first gravitational
wave detections~\cite{Abbott:2016blz,TheLIGOScientific:2017qsa}.  Ultra-high energy neutrinos ($>10^{17}$\,eV) will be unique 
messengers to the most
powerful astrophysical processes at cosmic
distances~\cite{Ackermann:2019ows}
and unique probes of
fundamental physics at extreme energies~\cite{Ackermann:2019cxh,Aartsen:2017kpd,Bustamante:2017xuy}.

Polar ice sheets are being utilized by many
experiments as a detection medium for 
high-energy astrophysical neutrinos~\cite{Aartsen:2016nxy,IceCube-Gen2:2020qha,Prohira:2021vvn}, including
many that are designed to 
detect neutrinos via a broadband ``Askaryan''
radio impulse~\cite{PUEO:2020bnn,Gorham2009,Allison:2011wk,Barwick2015,Anker:2020lre,RNO-G:2020rmc}.  So
far no neutrinos have been detected with radio techniques.  Due to
the transparency of pure ice at radio 
frequencies, the neutrino-induced 
signals will have
propagation distances of order kilometers in the ice before being detected by antennas either from
within or above the ice.  Thus, it is important to understand the impact of the ice on
the properties of signals as they propagate 
over that distance scale.  Ice crystals are 
known to be birefringent and in some locations it
has been shown that
polar ice
can
be treated as biaxial at radio
frequencies~\cite{Matsuoka2009}. This comes about 
because there are two special
directions in the ice: the direction of ice flow, which is in the horizontal plane, and the vertical direction due to compression.


Previous studies have investigated 
properties of polar ice and their
impact on
the detection of neutrino-induced radio 
impulses using 
diverse datasets and detailed 
simulations.  Radio-frequency measurements 
in polar ice
have characterized the
depth-dependent indices of refraction
$n(z)$
and the attenuation of signal power at
radio frequencies
 ~\cite{Kravchenko2004,Barwick:2005zz,Allison:2011wk,Avva:2014ena}.
 That $n(z)$ is not constant can bring
 about two solutions for rays
 propagating between source
 and receiver called direct and reflected/refracted, and both have been
 observed.
Simulation studies have stressed 
the importance of 
using techniques beyond simple ray tracing
to model signal propagation, especially in
 non-uniform ice  ~\cite{Deaconu:2018bkf,RadarEchoTelescope:2020nhe}.
 In Ref.~\cite{Barwick:2018rsp}, 
 previously unanticipated modes of 
horizontal propagation in the ice were reported.
Ref.~\cite{Besson:2010ww} reported 
evidence of birefringence in the ice sheet
as seen in delays between signals of different polarizations
using bistatic radar from the surface.

\begin{figure}
    \includegraphics[width=0.49\textwidth]{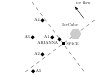} 
    \caption{Layout of the SPICE core, the five ARA stations,
    the ARIANNA South Pole station, and IceCube in the
    northing-easting coordinate system.  The arrow shows the
    direction of ice flow, and the dashed gray lines 
    intersect the location of the SPICE core and are
    parallel to and perpendicular to ice flow.}
    \label{fig:station_layout}
\end{figure}
The South Pole Ice Core Experiment (SPICE) calibration system
has provided a unique dataset to study
the impulsive signals with the transmitters, propagation  over
${\sim}{\rm km}$-scales,
and receivers all being within the
ice.
As seen in Fig.~\ref{fig:station_layout},
the ARA and ARIANNA in-ice neutrino 
experiments embedded in the ice 
measured radio-frequency impulses 
up to 4\,km
away from their source.
Five ARA stations (A1-A5) observed the pulses from
100 to 200\,m depths, and two 
ARIANNA stations from just below the surface.

In Ref.~\cite{Allison_2020}, using SPICE pulse measurements,
ARA reported
bounds on the depth-dependent 
index of refraction
at South Pole, time differences between
signals detected in different polarizations
due to birefringence, and attenuation lengths
for horizontally propagating signals.
ARIANNA reported measurements of the
polarization of signals from a SPICE
transmitter after propagating ${\sim}1$\,km distance in the ice~\cite{ARIANNA:2020zrg}.
Jordan {\it et al.}~\cite{Jordan:2019bqu} 
utilized
ice fabric properties from the
SPICE ice core to predict time
differences between signals in different 
polarizations that were
consistent with observations 
using a 
biaxial birefringence treatment for near-horizontal propagation and
restricted to special cases.  Besson {\it et al.}~\cite{Besson:2021wmj}
extended this work to vertical
propagation and additionally 
used past data
from the RICE experiment.

ARA and ARIANNA both observed
strange effects in the polarization of
SPICE pulses when viewing them at
different geometries~\cite{ARIANNA:2020zrg,Allison_2020}.
First, while SPICE pulses were transmitted predominantly in one polarization, they were often observed with larger-than-expected power in the other (cross-) polarization.  Sometimes even more  power was observed in the cross-polarization than in the transmitted polarization.
Second, the dependence of the observed signal
polarization on receiver positions and viewing angles
did not follow clear patterns.  
Similar effects have also been reported
in earlier measurements of radio signals
in polar ice after long propagation
distances in Ref.~\cite{Barrella:2010vs}.
In
this paper, I propose that 
birefringence that is effectively 
 biaxial at radio frequencies 
 in the ice is a plausible
explanation for these effects,  outlining a model to predict signal
polarizations with transmitter and
receiver both in the ice, and compare against
already reported observations of SPICE pulses.

Outside of radio neutrino detection, others have
investigated the impact of birefringence on signal
propagation in ice.  
Recently the IceCube
neutrino telescope at South Pole, 
which detects optical Cerenkov light from neutrinos
in the ice,  has reported
an anisotropic attenuation and attributes it to birefringence~\cite{Chirkin2019,tc-14-2537-2020}.  Since optical wavelengths
are much shorter than few-mm
individual
crystal sizes, they propagate
light through uniaxial ice crystals
whose orientations are distributed
as expected in the ice and that
have been elongated due to ice flow
and refract the light across
boundaries between crystals.
In recent years there has been much development in radar polarimetry,
which assesses the impact of the birefringence properties of the ice
on return power and polarization~\cite{Matsuoka2003,Matsuoka2009,tc-15-4117-2021,fujita_maeno_matsuoka_2006,Matsuoka2012,Young2021,Brisbourne2019,Jordan2019,Dall2010,jordan_schroeder_elsworth_siegfried_2020,tc-12-2689-2018,Yan} in radar measurements.

This paper is organized as follows.
First, I give a big picture
overview of the core concepts
behind this paper.  Then, I 
briefly summarize 
the already established theory behind 
the propagation of electromagnetic radiation
in biaxially birefringent crystals.  Next,
I describe the SPICE calibration
campaign and the ARA and ARIANNA
detectors at South Pole.  In the next
section, I
predict the behavior of received
signal spectra and power
measured in different polarizations
in the ARA and ARIANNA stations
after signal propagation while
treating the ice sheet as biaxially 
birefringent
and then compare 
with published results.
Finally, I conclude with implications
for experiments using radio techniques
to search for neutrinos in polar ice.

\section{Overview of the core concepts in this paper}

\begin{figure*}
\begin{tabular}[t]{|P{0.49\textwidth}|P{0.49\textwidth}|} 
\hline
 \includegraphics[width=0.35\textwidth,valign=T]{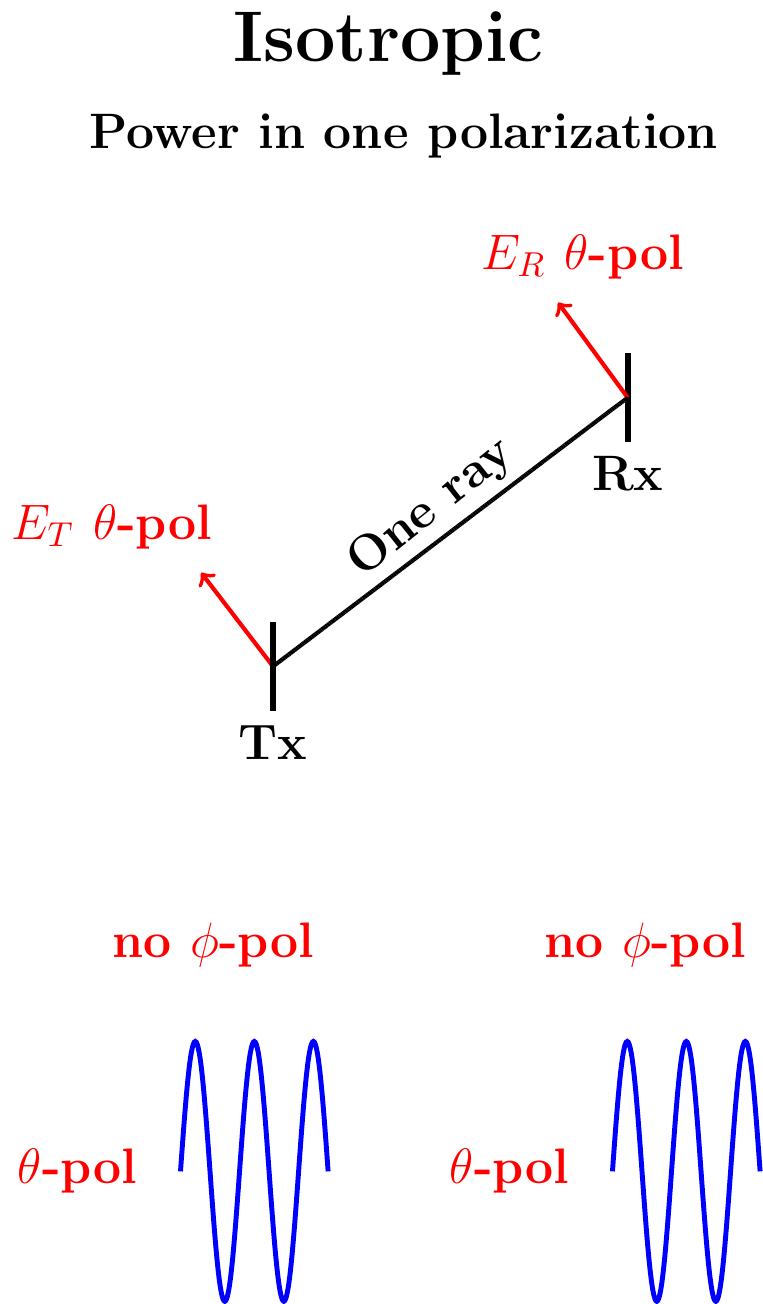} &  \includegraphics[width=0.385\textwidth,valign=T]{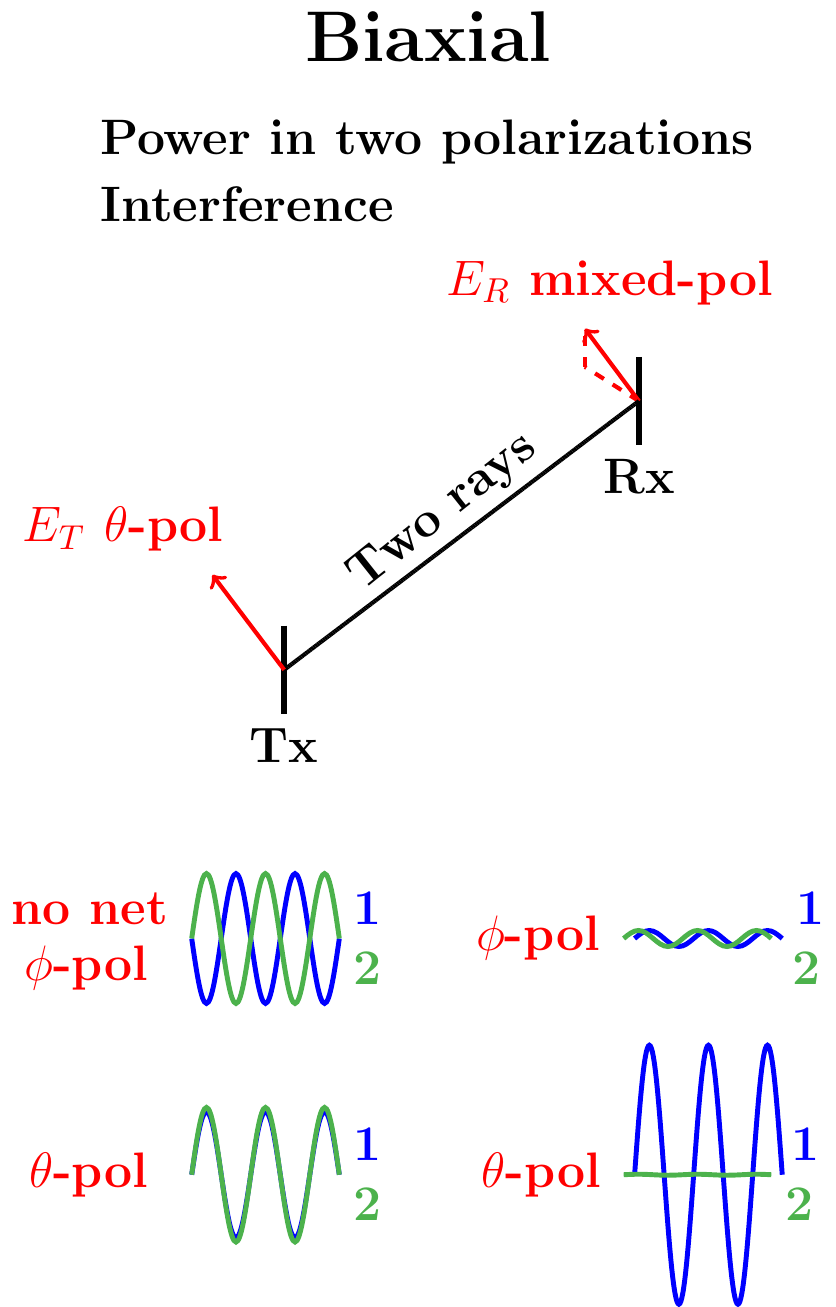} \\ 
  & \\
 \hline
\end{tabular}
\caption{\label{fig:illustration}Illustration
of a pure $\theta$-pol signal (no cross-pol) emitted by a transmitter (Tx) in the ice and propagated to
the position of a receiver (Rx) 
at a more shallow
depth.  Here, $\theta$-pol is the polarization in the plane
of the page and perpendicular to 
the ray, and 
$\phi$-pol is into the page.  Although this 
figure is only meant to serve as an illustration, it is modeled after a 300\,MHz
signal transmitted from the SPICE pulser at
1000\,m and received by an A1 station at 100\,m depth (more details of the stations in Sec.~\ref{sec:setup}).
(Left) In the case of an isotropic medium, there is one
direct ray and a pure
$\theta$-pol signal remains purely
$\theta$-pol at the receiver.  
(Right)
In a birefringent medium, there are
two direct rays\footnote{We will see that these two rays are a pair of eigenstate solutions to the wave equation.}.  Since here the signal
is purely $\theta$-pol, the $\phi$-pol components are such that they cancel each other.
During propagation, in a biaxially
birefringent medium,
each ray's envelope power 
can have its mixture of polarizations
altered, and the two rays can also interfere.
Both effects contribute to a mixed
 polarization of the signal at the receiver.
}
\end{figure*}
Fig.~\ref{fig:illustration} illustrates 
the main points underlying what I present 
in this paper.
I address, for the first time,
 the impact 
that birefringence that is effectively
biaxial will have on
the power received by
antennas measuring 
different polarizations 
in in-ice radio neutrino
experiments with the transmitter-receiver system
fully embedded in the ice.
I compare
expectations with previously reported 
measurements of SPICE pulses. 

In an isotropic medium (left panel of Fig.~\ref{fig:illustration}), a signal transmitted in one polarization is received in the same polarization and there is one 
direct path taken by a single ray between transmitter and receiver\footnote{Here I refer to the direct ray as opposed to the reflected or refracted ray, which can be present for a depth-dependent index of refraction.}.
Throughout this paper, a ``ray'' 
is defined by a path running tangent 
to its wave
vector ${\bf k}$ and has a unique displacement vector 
${\bf D}$ and electric field ${\bf E}$
associated with it at any given time.
Also, $\theta$-pol is the polarization in the plane
of the page and perpendicular to ${\bf k}$, and 
$\phi$-pol is into the page.

The right panel of Fig.~\ref{fig:illustration}
illustrates the effects considered
in this paper that 
biaxial birefringence can have
on the polarization of signals.
I consider two rays along the same direct
path (always with the same ${\bf k}$ between
them), but each ray
has its own ${\bf D}$ and electric field ${\bf E}$\footnote{In biaxial birefringence there
will be more than one direct ray path 
due to different polarizations seeing
different index of refraction profiles,
but in this paper, I only consider one direct
path.}. 
Here the signal is transmitted in
the $\theta$-pol polarization in a
birefringent
medium that is effectively biaxial, and so the $\phi$-pol components from
the two rays cancel one another.

Two different effects can cause
the polarization  
observed at the
receiver to be different from the one
transmitted as seen in Fig.~\ref{fig:illustration}.
First, the polarization of each ray
will change along its path for two reasons: 
the properties of the birefringent crystal 
are
depth-dependent, and the ray's direction (the direction of ${\bf k}$) 
changes 
with respect to the crystal axes due to the
depth-dependent index of refraction. 
This is illustrated in Fig.~\ref{fig:illustration} as the same ray
having different polarization components
at the transmitter and receiver.
Second,
in each polarization, the electric
fields associated with the 
two rays will interfere. 
This is illustrated as the two rays 
having a relative delay at the receiver
in either polarization. 
Both of
these effects will impact the polarization of the signal
detected at the receivers.   
I note that in a birefringent medium, two rays
may further split along their path into
more rays, but I neglect this and only consider
the two rays that leave the transmitter.

In this paper, I lay out 
the effect of a biaxial treatment of birefringence
on long-distance propagation with the 
source and transmitter both in the ice, and
to this aim, I make a few simplifications.  First, only a single frequency is ever propagated at a time.
Second, while I use a depth-dependent
index of refraction, I only consider the direct
solution between the transmitter and receiver
and neglect the reflected/refracted solution
that is often present.
Third, I consider only two rays, without
allowing for any further splitting of rays along
the path.  This is justified by the observation
of SPICE pulses that remain impulsive at the receivers with a single time delay between pulses in the two polarizations~\cite{Allison_2020}.
Fourth, I take the power in each ray
to be constant along their path.  So, the power
in ray 1 at the transmitter stays in ray 1
even as the polarization of ray 1 rotates along the path, and the same for ray 2.  This appears to be also the approach taken in Ref.~\cite{Matsuoka2009}.
Finally, I take the direct path to be 
the same for the two rays (both rays will 
have the same ${\bf k}$ at any time), while other
properties of the ray (${\bf D}$, ${\bf E}$,
and Poynting vector ${\bf S}$) will all be different between the two rays.  It will
be important to consider the 
more general cases in future work.

The formalism presented here for
signal propagation in ice 
using a biaxial
treatment of birefringence
appears to be consistent
with the one laid out for radar returns
at oblique angles in Matsuoka {\it et al.}~\cite{Matsuoka2009}, and the one
in Fujita {\it et al.}~\cite{fujita_maeno_matsuoka_2006} 
for vertical propagation.
However, unlike both of those, 
I do not include any loss of
power due to scattering.  

Scattering would only have an important
impact on this paper if it were anisotropic,
and only if it is large enough to impact
the transmitted power.
Ref.~\cite{Matsuoka2009} reports that
from radar data in Antarctica,
anisotropies in backscattered power can at most
change the return power by
about 10-15\,dB~\cite{Matsuoka2003,Matsuoka2004}, 
and 
so for a typical return
power of -80 to -60\,dB, it only affects
the transmitted power at typically at the -60\,dB level (one part
in a million).

\section{\label{sec:background}Background on birefringence}

A birefringent crystal is an anisotropic
medium, meaning that the propagation of electromagnetic
radiation depends on its direction 
and polarization due
to a distinctive feature of 
one or more axes within the crystal.  
\footnote{Note that anisotropy is distinct from 
whether a medium is inhomogeneous, 
 which is where the propagation
of EM radiation depends on position.
For example, a medium with a
depth-dependent
index of refraction
is inhomogeneous but in principle 
may or may not be
isotropic.}
An anisotropic medium may exhibit a
symmetry about one axis, in which case
it is uniaxially birefringent and is
characterized by two parameters. Biaxial
birefringent crystals are characterized by
three parameters along three perpendicular
axes.

While individual ice crystals are
uniaxially birefringent, at radio
frequencies the electromagnetic
properties of the ice depend on both
the birefringence of the individual
crystals and the distribution of
crystal orientations, known as the
crystal orientation fabric (COF) within the ice
volume (the bulk) ~\cite{Hargreaves1978}.  At radio
frequencies, the $\mathcal{O}$(1\,m) wavelengths
are much larger than the typical 
$\mathcal{O}$({\rm a few}\,mm)-sized crystals in polar ice~\cite{Alley2021}.

Ice Ih is the form of ice found in
ordinary water that has been
frozen at atmospheric pressure, or has
been formed directly from water vapor at
$>100\,^{o}$\,C~\cite{PetrenkoWhitworth}.  An Ice Ih crystal
consists of stacked planes of H$_{2}$O molecules that form 
a hexagonal structure in each plane, with the
hexagons in neighboring planes aligned.  The 
oxygens  sit at the vertices of
hexagons, and hydrogen forms
bonds with neighboring oxygens to connect
the lattice.

The hexagonal structure of 
Ice Ih crystals
leads to the axial symmetry that gives rise
to uniaxial birefringence.
Here, 
as in~\cite{Matsuoka2009}, I model the ice
as a biaxially birefringent
crystal even though it is composed
of uniaxially birefringent crystals.  
This is motivated by
the crystal orientations being influenced
by two special axes:
 1) the vertical axis, due to
compression and 2) the direction
of the flow of ice, which is in the horizontal
plane.  

In this section, I provide
the reader with the basics behind
electromagnetic waves in biaxially 
birefringent media and then outline
how I apply that theory to the case
of transmitting and receiving radio frequency (RF) signals
in South Pole ice.  I refer the reader
to Refs.~\cite{BERRY200713,Sjoberg,Keller_Lecture3,Orfanidis2013,StoiberMorse,Nelson,Druschel,Kuzel,SalehTeich} for more complete
treatments.
There are many places where I refer to a vector
without the hat, even when only the direction is needed,
to simplify notation.  I do use the hat when it is necessary for units.

\subsection{Electromagnetism in biaxially
birefringent crystals}
In a biaxially birefringent crystal,
for a given wave vector ${\bf k}$, 
there are
 not one but two ``rays,'' and each 
 propagates with a different
 index of refraction.  Those 
 two rays 
 have displacement vectors
 ${\bf D_1}$ and ${\bf D_2}$ that are perpendicular
to one another.
 As in isotropic media, 
the displacement vectors
sit in the plane perpendicular 
to ${\bf k}$.

\begin{figure*}
\centering
    \includegraphics[width=0.49\textwidth]{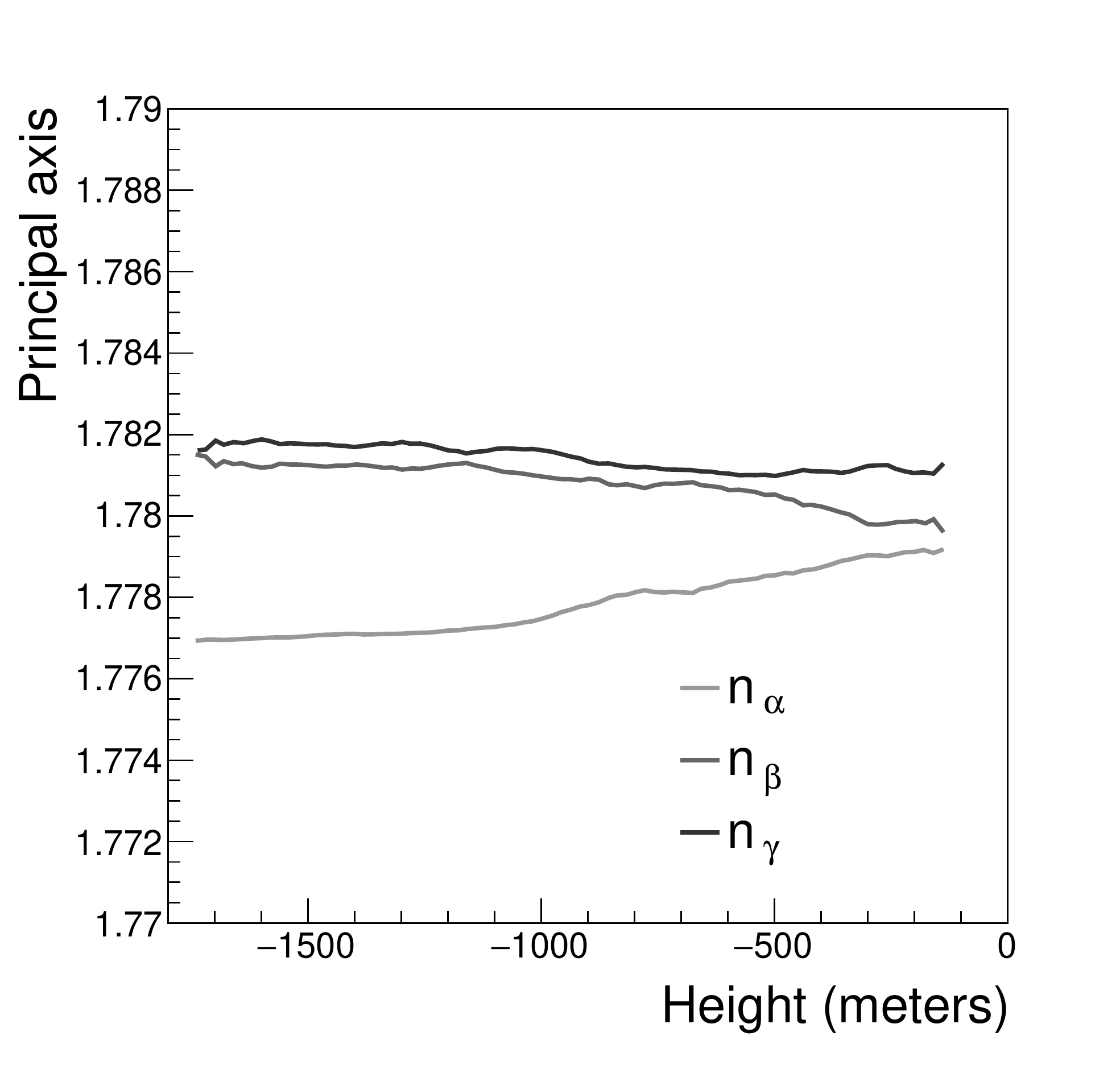}
    \hspace{0.5in}
    \includegraphics[width=0.35\textwidth]{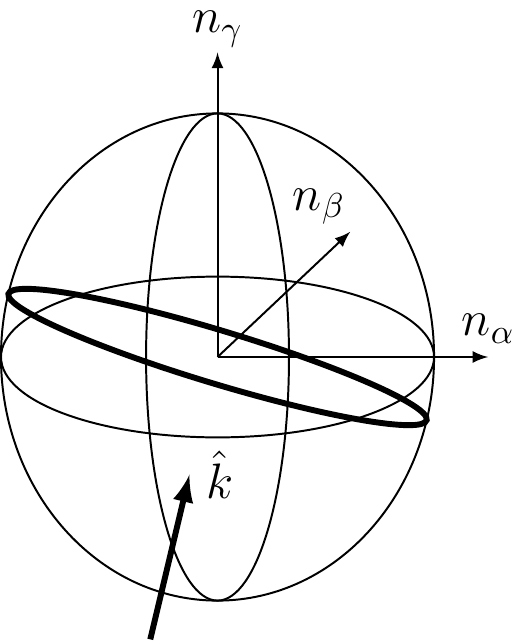}
    \caption{(Left) Principal indices of refraction used in this paper as reported by Voigt~\cite{Voigt:2017}
    and also utilized in~\cite{Jordan:2019bqu}.
    For the work in this paper, I smooth the data 
    with a three-point running average.
     (Right) Illustration of an indicatrix
  with dimensions exaggerated compared to what
  is observed in South Pole ice.
  An ellipse is shown that represents the 
  intersection of a planar wavefront
  for an incident wave vector ${\bf k}$
  with the indicatrix.  The eigenvectors of
  the displacement vector ${\bf D}$ are
  the major and minor axes of that 
  ellipse of intersection.   }
    \label{fig:n123}
\end{figure*}

To find 
${\bf D_1}$ and ${\bf D_2}$
for a given ${\bf k}$,
it is useful to think of the
wavefront encountering three
parameters $n_\alpha$,
$n_\beta$, and $n_\gamma$, which are
properties of the medium,
and at South Pole
 are depth-dependent (see Fig.~\ref{fig:n123}).  
These parameters form
the three perpendicular semi-axes of
an ellipsoid
called an ``indicatrix,''
and here the $\alpha$- and $\gamma$-axes 
are taken to be aligned with ice flow and 
the compression, respectively.  
The semi-axes are also called principal
axes.
Although they are indices of refraction, we will see that a ray only propagates with its
indices of refraction equal to 
$n_{\alpha}$, $n_{\beta}$, or $n_{\gamma}$   in special cases.
For an isotropic medium, $n_\alpha=n_\beta=n_\gamma$,
for a uniaxially birefringent medium 
two semi-axes are the same, and for a biaxially
birefringent medium such as we treat South Pole ice
in this paper
they are all different, and by convention,
$n_\alpha<n_\beta<n_\gamma$.  

Now we can find the directions of  
${\bf D_1}$ and ${\bf D_2}$
given ${\bf k}$, $n_\alpha$,
$n_\beta$, and $n_\gamma$.  
The
intersection of the
planar wavefront, which is perpendicular
to ${\bf k}$, with the 
indicatrix forms an ellipse.
The directions of  
${\bf D_1}$ and ${\bf D_2}$
are then in the directions of the
 major and minor axes of
that ellipse, and the lengths
of the major and minor axes are the 
indices of refraction seen 
by the two rays.

The direction of
the electric fields ${\bf E_1}$ and ${\bf E_2}$
associated with the two rays
are related to ${\bf D_1}$ and ${\bf D_2}$
using an equation of a familiar
form:
\begin{equation}
    {\bf D_{1,2}}= {\bm \varepsilon} {\bf E_{1,2}}.
    \label{eq:DequalsepsilonE}
\end{equation}
While in
isotropic media they would be related by 
a scalar $\varepsilon$, in birefringent media, ${\bm \varepsilon}$
is a tensor given by:
\begin{equation}
\begin{bmatrix}
n_\alpha^2 & 0 & 0 \\
0 & n_\beta^2 & 0 \\
0 & 0 & n_\gamma^2 \\
\end{bmatrix}.
\end{equation}
This means that
${\bf D}$ is not in general parallel to 
${\bf E}$, though at South Pole they are within a fraction of a
degree.  Recalling that ${\bf k}$
is perpendicular to ${\bf D}$,
then ${\bf k}$ is in general not perpendicular
to ${\bf E}$.  The electric fields ${\bf E_1}$
and ${\bf E_2}$ are perpendicular to
one another.

Considering the first ray, ${\bf k}$,
${\bf E_1}$, and ${\bf D_1}$ are
all related by the wave equation,
which now has an extra term compared
to the more familiar form due
to ${\bf E}$ not being perpendicular to ${\bf k}$:
\begin{equation}
    (-{\bf k}\cdot{\bf k}){\bf E_1} + ({\bf k}\cdot{\bf E_1}) {\bf k} = -\mu_0 \omega^2 {\bm \varepsilon} {\bf E_1}.
    \label{eq:waveequation}
\end{equation}
Recall that on the right side
of the equation, ${\bm \varepsilon} {\bf E_1}={\bf D_1}$.
Here, $\mu_0$ is the usual
permeability constant and
$\omega$ is the angular frequency 
of the wave.
This equation of course also must
be satisfied with
${\bf E_1}\rightarrow {\bf E_2}$, and ${\bf D_1}\rightarrow {\bf D_2}$.
The displacement vectors
${\bf D_1}$ and ${\bf D_2}$
corresponding
to a given ${\bf k}$ are the
two eigenvectors of this
wave equation and the 
indices
of refraction seen by each ray 
are the
corresponding eigenvalues.

In biaxially birefringent media,
the Poynting vector ${\bf S}$ points
in the direction of energy flow,
just as in isotropic media,
but now for a given ${\bf k}$, there are two Poynting vector directions, ${\bf S_1}$  and ${\bf S_2}$, one for each of the two rays, and neither is in general parallel
to ${\bf k}$.  These are found
from the usual relationship:
\begin{equation}
    {\bf S_{1,2}}={\bf E_{1,2}}\times{\bf H}.
    \label{eq:poynting}
\end{equation}
Here, the direction of ${\bf H}$ can be 
found by crossing ${\bf k}$ into ${\bf D}$:
\begin{equation}
    {\bf \hat{H}_{1,2}}={\bf \hat{k}_{1,2}} \times{\bf \hat{D}_{1,2}}.
    \label{eq:H}
\end{equation}

\subsection{The $\epsilon$ angles}
\label{sec:epsilons}

\begin{figure*}
   \begin{center}
\includegraphics[width=0.45\textwidth,valign=T]{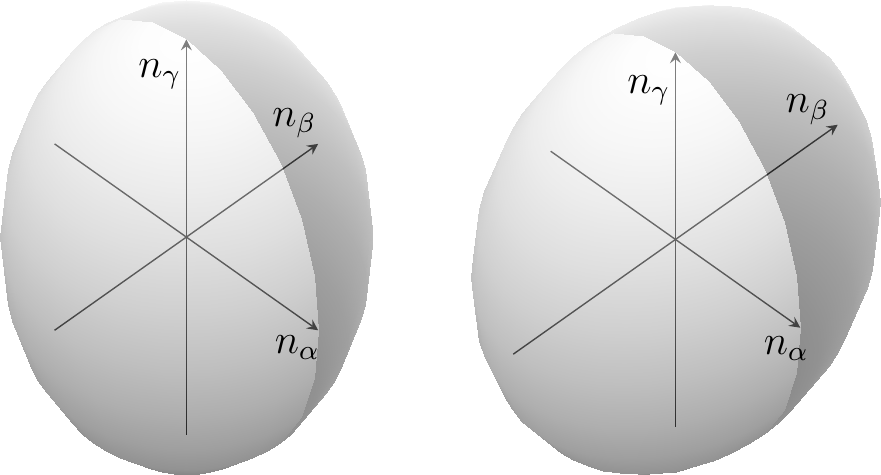}
\caption{\label{fig:epsilons} These two ellipsoids are not
the same and will be used in Fig.~\ref{fig:epsilons_intersection} to demonstrate the reason
that a biaxial treatment of 
birefringence is needed for the $\epsilon$ angles
to be non-zero.  
(Left) Indicatrix for a uniaxial crystal
(here, $n_{\alpha}=n_{\beta}=1.0$, $n_{\gamma}=1.5$), and
(Right) indicatrix for a biaxial crystal
(here, $n_{\alpha}=1.0$, $n_{\beta}=1.3$,
$n_{\gamma}=1.5$).  I take
indicatrices with features exaggerated
compared to those observed in the ice
for illustration purposes.}
\end{center}
\end{figure*}

\begin{figure*}
\includegraphics[width=0.95\textwidth]{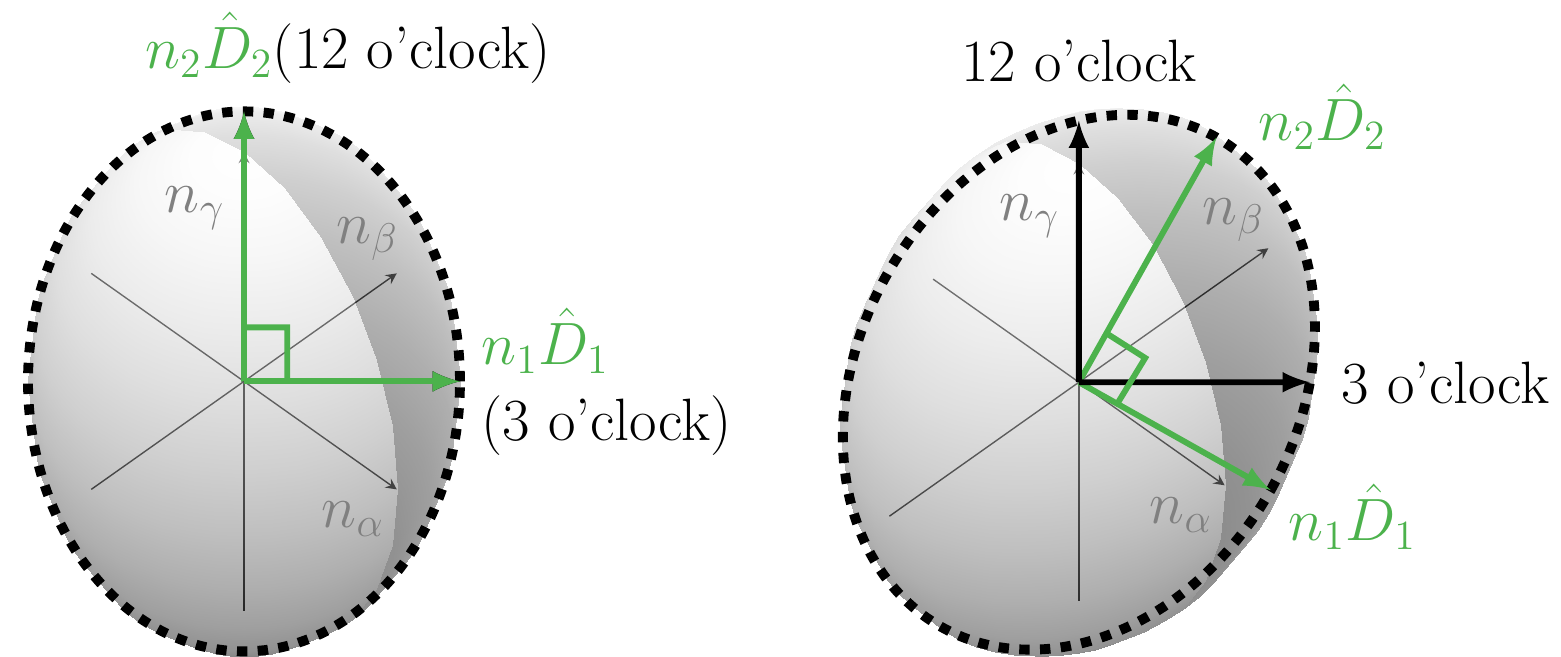}
\caption{\label{fig:epsilons_intersection} These are the same two ellipsoids as the ones
shown in Fig.~\ref{fig:epsilons}.  
(Left) Indicatrix for a uniaxial crystal, and
(Right) indicatrix for a biaxial crystal.
Consider ${\bf k}$ for an incident ray to
be into the page, where it hits
indicatrix at oblique angles (here, $\theta=45^{\circ}$, $\phi=45^{\circ}$
in a usual spherical coordinate system).
The planar wavefront is in the plane of the page.
The outline of the
shaded region shown as the dashed black line is the intersection ellipse, that is, the intersection of the planar wavefront
with the ellipsoid.
The axes of the ellipse are the eigenvectors of ${\bf D}$.
For the uniaxial crystal,
the eigenvectors are at 12 o'clock
and 3 o'clock as seen by an observer looking along ${\bf k}$, and for the biaxial crystal, 
the eigenvectors are rotated by an angle in the plane of the page
with respect to 12 o'clock and 3 o'clock.}
\end{figure*}

In this paper, I highlight two problems seen
in the data and describe how a
biaxial treatment of birefringence in the ice may explain them: 
1) detecting
power in $\phi$-type antennas
after transmitting from a $\theta$-type
antenna and 2) the complicated 
dependencies between the
relative power
in $\phi$-type and
$\theta$-type antennas and the positioning
of the transmitter relative to the
receiver.  I introduce angles
that I call $\epsilon$ that I
view as being at the core of the emergence of these effects.  The $\epsilon$ angles 
also point to a {\it biaxial} treatment of birefringence being needed to bring about the behaviors seen in the data,
as they would not be present in either a uniaxially birefringent or isotropic medium.
I note that $\epsilon$ is referred to
as $\varphi$ in Matsuoka {\it et al.}~\cite{Matsuoka2009}.

We will consider a biaxially birefringent
crystal and a uniaxial one, shown in Fig.~\ref{fig:epsilons}, and compare their effect on the eigenvectors of
${\bf D}$, and thus the orientation of
the electric fields.  For illustration
purposes, the differences between semi-axes
are exaggerated in this figure 
compared to what is expected in the ice.  In this section, for the uniaxial crystal
I take $n_{\alpha}=n_{\beta}=1.0$ and $n_{\gamma}=1.5$,
and for the biaxially birefringent crystal I take $n_{\alpha}=1.0$, $n_{\beta}=1.3$, and $n_{\gamma}=1.5$.

 Recall that 
 for a given ${\bf k}$, there are
two eigenvectors for ${\bf D}$.
The planar wavefront is perpendicular
to ${\bf k}$.
The directions of the eigenvectors
of ${\bf D}$
are
along the major and minor axes of
the ellipse defined by the intersection of
the planar wavefront and the indicatrix.

Fig.~\ref{fig:epsilons_intersection} illustrates the effect that these
two types of crystals have on the 
orientations of the eigenvectors of ${\bf D}$. On each side of the figure,
consider ${\bf k}$ to be 
approaching the indicatrix
into the page and perpendicular to the page.
On the left side of the figure, 
when ${\bf k}$ is incident on the
uniaxial indicatrix,
there is a symmetry about the $\gamma$-axis, and
to an observer looking in the direction of
${\bf k}$,
the axes of the intersection ellipse will
appear to sit at 12 o`clock and 3 o`clock.
What I call 12 o`clock is the direction that is perpendicular
to ${\bf k}$ and in the 
plane of ${\bf k}$ and the $\gamma$-axis.  On the right
side of the figure, when
${\bf k}$ is incident on a biaxial
crystal,
the symmetry is lost, and  the eigenvectors
of ${\bf D}$ will, in general, be rotated with respect
to 12 o`clock and 3 o`clock by an angle.

The angles that the ${\bf D}$ eigenvectors make
with the 12 o'clock and 3 o'clock directions
are 
almost what I call the $\epsilon$ angles but not quite, because polarizations are in
the directions of the corresponding
electric fields ${\bf E_{1,2}}$ for
the two eigensolutions.  Recall that 
${\bf E}$ and ${\bf D}$ are not in the
same direction although they are within a fraction of a 
degree of one another.  The electric fields are perpendicular to the
directions of the
Poynting vectors ${\bf S_1}$ and 
${\bf S_2}$,
not ${\bf k}$, and the latter three are not in the same
direction, but are also within a fraction of a degree
of one another (see Appendix~\ref{sec:k_S}).
I will define angles 
$\epsilon_{1}$ and $\epsilon_{2}$ 
 to be
the angles that the eigenvectors ${\bf E_1}$ 
and ${\bf E_2}$
make with respect to 12 o`clock and 3 o`clock
from the perspective of an observer looking 
in the directions of ${\bf S_1}$ and
${\bf S_2}$, respectively.  In each case, 12 o'clock
is perpendicular to ${\bf S}$ and 
in the plane containing ${\bf S}$ and 
the $\gamma$-axis. 
\begin{figure*}
    \centering
    \includegraphics[width=1.0\textwidth]{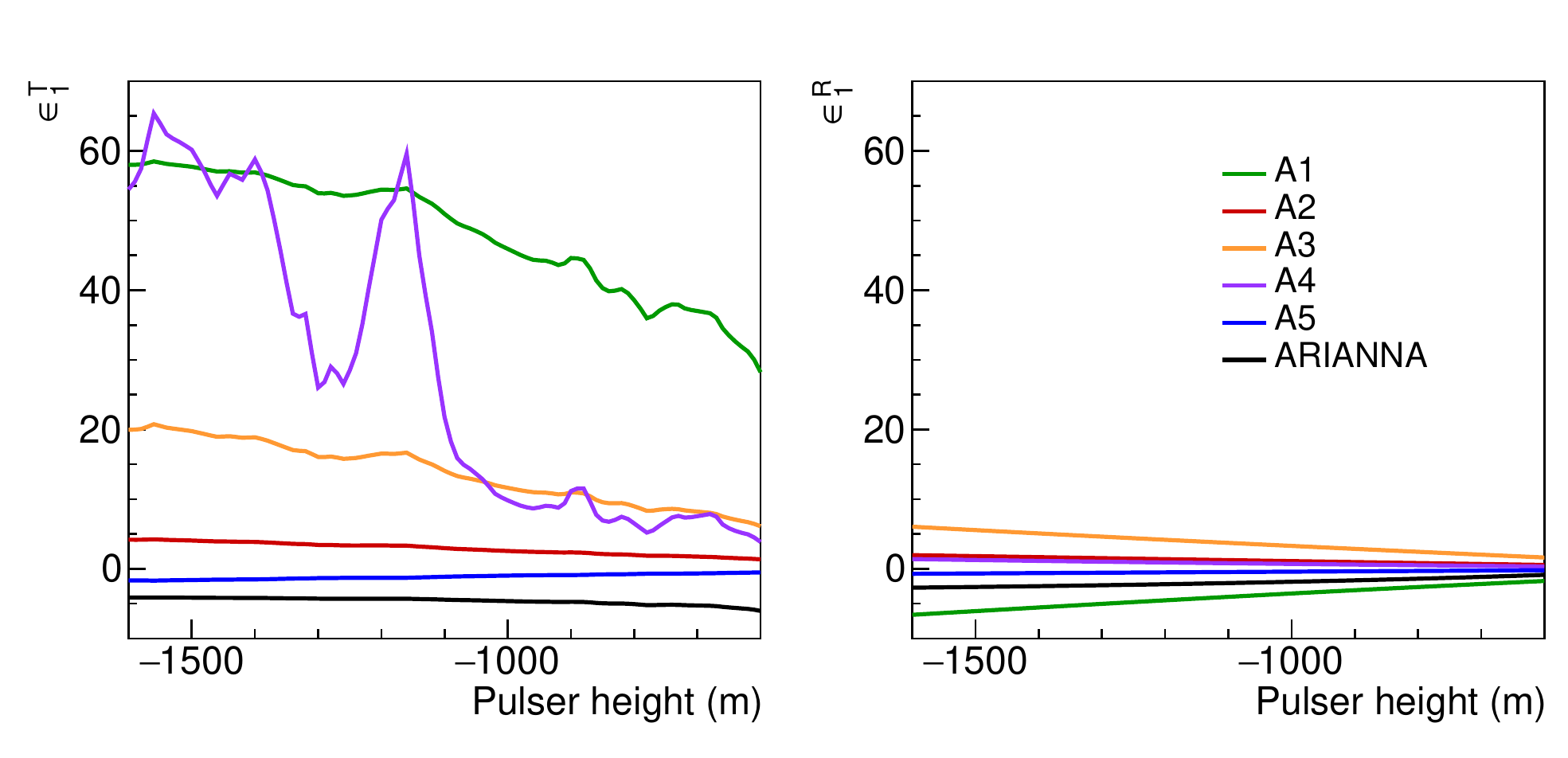}
    \caption{Angles that the electric fields
    make with the 12'oclock and 3 o'clock directions, called $\epsilon$ angles, for ray 1  emitted from the transmitter (left) and intersected by a 
    receiver (right) at each station as a function
    of pulser height.  The A1 antennas are at 80\,m depth, the antennas in other ARA stations at 180\,m depth, and ARIANNA antennas at the surface.  For an isotropic
    or uniaxially birefringent medium, these
    epsilon angles would all vanish.  The two rays 
    have similar $\epsilon$ angles ($\lesssim 0.1^{\circ}$ difference) as
    can be seen in Fig.~\ref{fig:diffepsilons} in Appendix~\ref{sec:differences}.  I note that these $\epsilon$ angles are not
    frequency-dependent.}
    \label{fig:epsilons_sidebyside}
\end{figure*}

There are cases where the $\epsilon$ angles
vanish.  As we have seen in Fig.~\ref{fig:epsilons_intersection},
for a uniaxial crystal the symmetry 
necessary to maintain the eigenvectors at
12 o'clock and 3 o'clock 
will be maintained
no matter the angle of approach.
In a biaxial crystal, the symmetry 
about one axis that keeps the $\epsilon$
angles at zero is in general only
maintained if ${\bf k}$ is in
a plane formed by two of the principal axes.  I note that in Jordan {\it et al.}~\cite{Jordan:2019bqu},
propagation was only carried out in the $\beta-\gamma$ 
and $\alpha-\gamma$ planes.  In a 
biaxially birefringent
medium, in the $\alpha-\gamma$ plane,
there is even one angle with respect to the 
$\gamma$-axis where the medium appears
 as if it were isotropic to an incident wave (see Appendix~\ref{sec:V}).

Fig.~\ref{fig:epsilons_sidebyside} shows the $\epsilon$ angles for ray\,1 seen by a SPICE
signal propagating from the transmitter to
the receiver
for the five ARA stations and ARIANNA.  The differences
between the $\epsilon$ angles for the two rays are small ($<0.1^{\circ}$ and are shown in Fig.~\ref{fig:diffepsilons} in Appendix~\ref{sec:differences}. Assuming the indicatrix
is not frequency-dependent, then the
epsilon angles are also not frequency-dependent.
Note that the epsilon angles are often non-zero and can be large, and can change greatly between the transmitter and receiver.  The non-zero epsilon angles 
will enable interference patterns at the receiver even
in the absence of cross-polarization at the transmitter.  That the epsilon angles change between the transmitter and receiver will cause
a signal transmitted purely in $\theta$-pol
to be detected with $\phi$-pol power.  See
Fig.~\ref{fig:epsilon_vs_k} in Sec.~\ref{fig:epsilon_vs_k}
for the dependence of $\epsilon$ angles
as a function of all directions of $\vec{k}$.

The $\epsilon$ angles can change between
the transmitter and receiver for two
reasons. First,
because the indicatrix is depth-dependent, and second
because the direction of ${\bf k}$ changes along
the path due to the depth-dependent indices
of refraction.  Both of these effects change the intersection
ellipse.

Which of these two effects, a depth-dependent indicatrix or a changing
${\bf k}$ along the ray,
is the dominant
reason for changing $\epsilon$ angles
appears to be geometry-dependent.  In
the case of A1, the ray path is nearly a 
straight line as illustrated in Fig.~\ref{fig:illustration}, yet the $\epsilon$
angles still change significantly, and so the
depth-dependence of the indicatrix must
be the dominant cause of the change in
that case. However, notice in Fig.~\ref{fig:epsilons_sidebyside}
the dramatic changes in
$\epsilon$ angles at the
transmitter depths between 
1500\,m and 1000\,m for signals destined for A4.  At the same depths, we can see in Fig.~\ref{fig:n123} that the indicatrix is only changing slowly.  The large
changes in $\epsilon$ angles seen by
signals between the transmitter and 
A4 must be dominated by the changing
direction of ${\bf k}$.
Characterizing the 
source-receiver geometries
that will see the greatest changes in
$\epsilon$ angles will be a topic of future study.

\section{Experimental setup}
\label{sec:setup}
The South Pole ice core (SPICE)
was a ${\sim}40$\,ka (40,000-year-old), 1500\,m deep
core of ice recovered in the 2014-15
and 2015-16 Austral summer seasons
for investigations of glaciology and climate 
history from the ice at South Pole~\cite{Casey2014}.
Subsequently,
from Dec. 23$^{\rm rd}$ to Dec. 31$^{\rm st}$ of 2018, members of the ARA and 
ARIANNA collaborations lowered
pulsers transmitting radio-frequency
pulses into the core so
that they would be observed by
the (five) ARA and (one) ARIANNA stations nearby~\cite{Allison_2020,ARIANNA:2020zrg}.  This provided an unprecedented
dataset for measuring signals after
propagating up to 4\,km horizontal 
distance with the
transmitter and receiver both in
the ice.  Measurements from
all six stations were made in different polarizations as a function of the depth of the pulser, from
the different vantage points and distances of the six receiver stations (see diagram in Fig.~\ref{fig:station_layout}).

The depth-dependent properties of the 
SPICE ice core 
were
reported in Ref.~\cite{Voigt:2017} and are summarized 
in 
Fig.~\ref{fig:n123}.  Samples of the 
ice from every 20\,m in the core,  from 140\,m to 1749\,m 
depth, were analyzed at Penn State University using
a c-axis fabric analyzer, which uses
cross-polarized light to find the 
average orientation of 
the c-axis, which is the direction of approach at which the crystal
behaves as an isotropic medium.
As a function of depth, the c-axis was found to rotate
within a vertical plane, becoming more vertical with increased
depth.  Although the orientation of the plane about the vertical
axis is not 
known since the orientation of the core was not preserved
during data taking, it is assumed
based on understanding that it is
aligned with 
the direction of ice flow, which is 36$^{\circ}46'23''$ counterclockwise from the northing direction in the northing-easting
coordinate system used by South Pole surveyors.

The SPUNK PVA (SPICE Pulser from UNiversity of Kansas Pressure Vessel Antenna) 
transmitter was an aluminum fat
dipole modeled after the ones
used for the RICE experiment~\cite{RICE:2001ayk},
described in detail in Ref.~\cite{Allison_2020}.
The $9$\,cm~$\times~90$\,cm antenna was
made to fit the 97\,mm hole and 
designed so that the impedance would
be 50\,$\Omega$ to match the cable 
when immersed in the estisol-240 drilling fluid environment in the hole.

ARA and ARIANNA are both neutrino experiments aiming
to detect ultra-high energy neutrinos above ${\sim}10^{17}$\,eV,
and stations from the different detectors,
ARA deep in the ice and ARIANNA at the surface,
provided views of the SPICE pulser from
many vantage points with respect to the principal
axes of what we are treating as a biaxial crystal from varying distances.
Table~\ref{tab:coordinates} gives the coordinates
of each station near South Pole and the 
angle that an observer sees
the pulser in the horizontal plane relative
to ice flow.

The Askaryan Radio Array (ARA) is a neutrino
detector at South Pole aiming to detect impulsive radio Askaryan 
emission from neutrino-induced cascades in the ice~\cite{Allison2016,Allison:2011wk,Allison2015,Allison2019,ARIANNA:2020zrg}.
ARA consists of five stations
of sixteen antennas deployed up to 180\,m deep in
the ice. 
Each station includes sixteen antennas, eight ``VPol''
and eight ``HPol'' (in this paper referred to
as $\theta$-pol and $\phi$-pol) with bandwidths spanning
150-800\,MHz.  The antennas of the former type are dipoles 
and the latter are ferrite-loaded quad slot
antennas.
The antennas in a station are arranged in approximately a 20\,m$\times$20\,m$\times$20\,m square
sitting along four vertical
strings, 
each with two  $\theta$-pol, $\phi$-pol
pairs 
of
top and bottom antennas.
When a 3/8 coincidence trigger in either polarization is satisfied, signals
are transmitted
to the surface where waveforms from all sixteen
antennas are read out in approximately 1000\,ns waveforms
at a sampling rate of 3.2\,GHz.

ARIANNA~\cite{Barwick2015,Barwick:2014boa,Anker2019,Barwick2006,Barwick2006arianna} is a neutrino detector aiming 
to detect the same Askaryan signature from neutrino-induced
cascades in the ice, with nine stations at
Moore's Bay
on the Ross Ice Shelf near the coast of Antarctica.   ARIANNA deploys log-periodic dipole antennas (LPDAs) at the surface, and being on the ice shelf it is sensitive to neutrino-induced 
radio emission reflected from the ice-water boundary below.  

ARIANNA deployed
an additional two stations  
near South Pole, and in Ref.~\cite{ARIANNA:2020zrg} reports
 results from observing SPICE pulsers in 
 what they call Station~51.  Station 51 consisted
of eight antennas, four of which are
down-facing LPDAs arranged in a 6\,m$\times$6\,m
square and oriented to measure two 
horizontal
polarizations perpendicular to one another 
in the plane of the square, at 0.5\,m below the surface.  There are
additionally four bicone antennas oriented
vertically at the corners
of the square.  Thus, the station measures
polarizations in three mutually 
perpendicular directions.
The ARIANNA station is
sensitive in the 80\,MHz-300\,MHz band and
digitizes signals at 1\,GHz.

\begin{table}[]
    \centering
    \begin{tabular}{c c c c p{1cm} p{1cm} }
    Station  & E (m) & N (m) & Distance (m) & Depth (m) & $\angle$ ($^{\circ}$)\\ \hline
    SPICE  & 12911 &	14927.3  & N/A & 600-1600 & N/A \\ 
       A1  & 11812.2	& 15560.3 & 1268 & 80 & 23\\
        A2 & 10814.6	& 13828.5 & 2367 & 180 & 81\\
        A3 & 9814.56& 	15561 & 3160 & 180& 42\\
        A4 & 10813.7& 	17293.4 & 3161 & 180& 4.8\\
        A5 & 9862.11&	12114.6 & 4148 & 180& 96\\
        ARIANNA & 12543.4	& 15356.4 & 565 & 1& 3.8 \\
    \end{tabular}
    \caption{\label{tab:coordinates}
    Coordinates of the SPICE ice core hole and 
    each station in Easting and Northing,
    the horizontal distance between the pulser
    and each station,
    depth of pulser/receivers, and the angle $\alpha$ that each station views
    the pulser in the horizontal plane relative
    to the direction of ice flow.  The ARA station positions
    are from~\cite{aracomm} and the ARIANNA
    station position is from~\cite{Geoff_github}.}
    \label{tab:my_label}
\end{table}

\section{\label{sec:pulsers}Pulsers in a Biaxially Birefringent Medium}

In this section, I describe the strategy that
I use to model 
the transmission 
of impulses by an antenna embedded
in an anisotropic medium
and their propagation in the ice.~\footnote{ \url{https://github.com/osu-particle-astrophysics/birefringence}.}
At the transmitter, the electric field of the signal will be written as the sum of the fields from two eigenstates.  Along the path, the two eigenstates
will follow the same path, but accumulate different phases, and their polarizations will rotate as
the eigensolutions for the electric field rotate.
While propagation of waves in
biaxially birefringent media is well
documented, it is not typical for
the transmitter itself to sit
within a dense medium, far less
an anisotropic medium.  Therefore, 
there are places where I need to
simply propose a way to proceed.

\subsection{Assumptions}
\label{sec:assumptions}

I consider two different
types of antennas, one of which is traditionally called VPol and
the other HPol, but for greater 
clarity here I call
them $\theta$-type
and $\phi$-type, respectively, because
they measure the ${\bf \hat{\bf \theta}}$ and
${\bf \hat{\bm \phi}}$ components of an incident 
field.  The physical
antennas of either type sit with
their orientation vertical (down a hole).
Where effective heights are needed, I use
simple forms of those for ARA 
antennas given by: ${\bf h^{\theta}_{\rm eff}}=h_0 \sin{\theta}\, {\bf \hat{{\bm \theta}}}$ and 
${\bf h^{\phi}_{\rm eff}}=h_0 \sin{\theta}\, {\bf \hat{{\bm \phi}}}$ for $\theta$-type and
$\phi$-type antennas, respectively.  
The effective heights will in general be
frequency-dependent, but
I only consider a single frequency at
a time in this paper.

The electric field of the signal emitted by
the transmitter is related to the
transmitted power and the effective height
${\bf {h}^T_{\rm eff}}$ of the antenna.  We
express the field at the transmitter as:
\begin{equation}
{\bf E_T^{\rm total}} = E_0 \, {\bf h^T_{\rm eff}}( \theta, \phi)  /h_0
    \label{eq:effectiveheight}
\end{equation}
where 
$\theta$ and $\phi$
are the zenith and azimuth of the direction
of the
ray at the transmitter, and the square
of $E_0$ is proportional to the transmitted power.

I propose that in an anisotropic
medium the $\theta$ and $\phi$ in Eq.(~\ref{eq:effectiveheight})
 are the angles
of ${\bf S}$ (not ${\bf k}$) and that
while ${\bf E}$ and ${\bf D}$ 
are not in the same direction,
it is still the electric field that is
dotted into the effective height (not ${\bf D}$).
Recall that ${\bf S}$ is in the direction
of energy flow and ${\bf E}$ is perpendicular
to ${\bf S}$.  
\footnote{Thank you to Patrick Allison, Jim Beatty, and Steven Prohira
for discussion on this point.}
Fig.~\ref{fig:k_S} in Appendix~\ref{sec:differences} shows the
angle between ${\bf k}$ and ${\bf S_1}$
and ${\bf S_2}$ for the SPICE pulses
viewed by the stations.  They differ at
most by about 0.2$^{\circ}$, which is approximately the angular resolutions of
the detectors on the directions of received signals.  After the development
of a more complete model for ray tracing
in biaxial birefringence, the choice of
${\bf S}$ here for the vector that defines
$\theta$ and $\phi$ could be experimentally
tested.

It is ${\bf k}$ that satisfies Snell's law,
and so a receiver will observe a ray if its
wave vector ${\bf k}$ takes a path that intersects both the transmitter and
receiver in the depth-dependent indices of refraction seen by
that ray~\cite{Keller_Lecture3,ChewJiao_Lecture17}.
The reason for it being ${\bf k}$ and not ${\bf S}$
that needs to satisfy this requirement is that 
the phase matching condition, which leads to Snell's law, requires that at an interface:
\begin{equation}
    {\bf k_{\rm inc}} \cdot {\bf x} = {\bf k_{\rm refl}} \cdot {\bf x} = {\bf k_{\rm trans}} \cdot {\bf x}
\end{equation}
where ${\bf x}$ is a vector parallel to the surface and pointing to a
location on the interface surface where
the wave vector is incident.  The subscripts denote
incident, reflected, and transmitted.
I call the direction of the
energy flow of the ray leaving
the transmitter ${\bf S_T}$.

\subsection{Ray Tracing}

\label{sec:raytracing}
Although the treatment of Snell's law in uniaxially birefringent media is common in the literature~\cite{Orfanidis2013},
the bending of rays in biaxially birefringent
media is rarely discussed
due to its complexity.  In Ref.~\cite{Latorre2012},
the authors lay out an iterative
procedure to find the ray solutions, and 
preliminary work on this problem
for in-ice neutrino detectors was presented in Ref.~\cite{Harty2021}.

In this paper, I simply take the rays
to bend as they would in an isotropic
medium with a depth-dependent index
of refraction
and
leave a more complete treatment for future work.  I also use the same $n(z)$ for both rays leaving the transmitter for finding the ray paths.  Just
as in an isotropic medium, I take the
rays to follow a path in the plane of the transmitter, receiver, and the vertical axis.  
Although the depth-dependent
index of refraction does lead to both ``direct'' and
``refracted'' solutions, the latter reaching the receiver
after a downward bend, for simplicity I only consider the
direct ray solutions here.

Fig.~\ref{fig:epsilon_vs_k} in Appendix~\ref{sec:epsilon_vs_k} allows us to evaluate
the potential 
impact of using ray tracing in an isotropic medium with a simple $n(z)$ 
even though 
the fields are propagated in a biaxial birefringent medium
described by $n_{\alpha}$, $n_{\beta}$, and $n_{\gamma}$.
Measurements of the arrival directions of 
SPICE pulses in ARIANNA have been consistent with those expected
from ray tracing in an isotropic medium
to within about $1^{\circ}$ in both zenith and azimuth~\cite{Anker:2020lre}, and
ARA reports SPICE pulse arrival directions to within a few degrees in zenith in A2~\cite{Allison_2020}.  Additionally, two 
calibration pulsers
were deployed along an IceCube string in that detector's final
season of construction at a distance of 4\,km, and both A2 and
A3 observed arrival directions of those pulses with a few degrees
in zenith and about $1-2^{\circ}$ in azimuth~\cite{Allison2016}. 
From Fig.~\ref{fig:epsilon_vs_k}, a deviation of the direction
of ${\bf k}$ by of order a degree will not have a qualitative
effect on the behavior of the $\epsilon$ angles, and thus the
rotation of the polarization vectors.

In Ref.~\cite{Latif2020},
the ray solutions are found 
for $n(z)$ profiles with the 
exponential form $n(z)=1.78-0.43e^{0.0132z}$ (where $z$ is negative and in meters),
but I use a modified form.
This is because the eigenvalues for the index of refraction
seen by any given wavefront will lie
between $n_\alpha$ and $n_\gamma$.  Therefore, 
I alter the exponential
parameter in the expression from Ref.~\cite{Latif2020} so that
the index of refraction is between $n_\alpha$ and $n_\gamma$ 
at depths where they are measured while keeping $n=1.35$ at the surface and $n=1.78 $ in deep ice.  I instead use the profile
$n(z)=1.78-0.43e^{0.03624z}$.  At $150$\,m depth, this
shifts the index of refraction from $n=1.721$ to 1.778.

Thus the ray tracing algorithm used
here is simplistic for many reasons.
First, the n(z) profiles seen by each 
ray should be different.  Second,
 in a biaxially birefringent medium,
a ray will not in general stay in the plane
we would expect in an isotropic medium.  Third, 
as the indicatrix changes along its path, 
a ray would continue to split into many rays
(because
the incident ray solution is not an eigenvalue of the new indicatrix at each step), but 
I neglect this effect.  This
is an area that requires much additional
effort.

I use a depth-dependent attenuation of
the field strength that is used by Ref.~\cite{Latif2020}, which comes from Ref.~\cite{ice_properties}.
Thus I apply an attenuation factor
$\mathcal{A}$ to account for the attenuation
of the electric field along the path of the 
ray
from the transmitter $T$ and
the receiver $R$
 given by:
\begin{equation}
    \mathcal{A}=\int_{T}^{R}{e^{-ds/\ell } ds}
\end{equation}
where $\ell$ is the field attenuation
length at the ice depth at position $s$
along the ray's path.

\subsection{Wave propagation}
\label{sec:propagation}
In this section, I lay out the procedure I
use to propagate signals from the transmitter to the receiver.
Throughout, I develop the mathematical
expressions for a single frequency.
Impulses would of course be a sum of
contributions of different frequency components
with appropriate phases.

\subsubsection{Simplification: Two Rays, Same Wave Vectors}
Let's begin with an approximation that
for a given transmitter-receiver pair,
there is only one ${\bf k}$
at the transmitter whose path
in the ice
will intersect the receiver.  For that
single ${\bf k}$ at the transmitter, there are two eigensolutions for ${\bf E}$ at the transmitter
that are perpendicular to each other, and
each sees a different index of refraction.
I keep these as two rays that propagate at different speeds with
 fields
${\bf E_{\rm 1}}$
and ${\bf E_{\rm 2}}$, but always with their
wave vectors in
the same direction at a given depth. 

The two rays 
will accumulate a phase difference due 
to seeing different indices of
refraction along their path.  My
simplification, however, forces
both rays to take the same path.
So, the phase differences that we will find only
come about due to the different
polarizations seeing different indices of refraction on their path, not
from any differences between their
path lengths.
For a 1\,km path length, a deviation of the path trajectory by $\sim1^{\circ}$ would lead to a path difference of approximately 1\,km $\cdot\cos{1^{\circ}}=15$\,cm, which at 300\,MHz is half a wavelength in deep ice for $n=1.79$, so a proper treatment of ray tracing could impact the positions
of the interference peaks and nulls.

\subsubsection{Propagating Signals from the Transmitter to the Receiver}

The signal propagation starts with
finding the two eigenvectors of the electric field at the transmitter
${\bf E_{1}^T}$
and ${\bf E_{2}^T}$ given ${\bf k}$
and the indicatrix at the transmitter depth.
To find these directions, 
we first
find the eigenvectors
of the displacement vector ${\bf D_{1}^T}$ and ${\bf D_{2}^T}$ corresponding
to ${\bf k}$ by finding the intersection of the planar wavefront with the indicatrix.
For this, I use the procedure laid out in Ref.~\cite{Bektas2016} and the
depth-dependent principal axes
shown in Fig.~\ref{fig:n123} from~\cite{Voigt:2017}.
The electric fields are then found from
${\bf E_{1}^T}={\bm \varepsilon}^{-1} {\bf D_{2}^T}$ and
${\bf E_{2}^T}={\bm \varepsilon}^{-1}{\bf D_{2}^T}$.

\begin{figure*}
\includegraphics[width=0.95\textwidth]{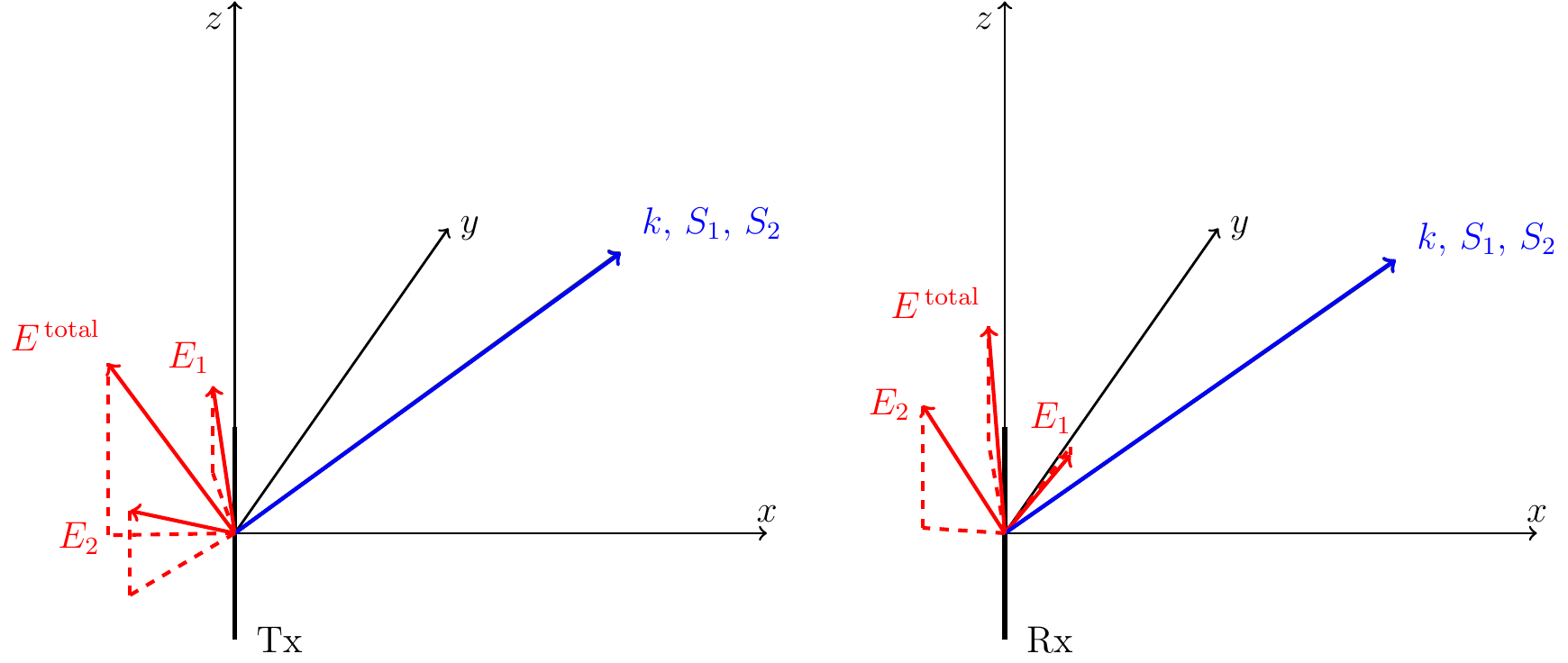}
\caption{\label{fig:vectors}Vectors associated with the two rays
at the transmitter (left) and the receiver (right)
when the transmitter is at 1000\,m depth and
the receiver at station A1. The differences between the ${\bf k}$, 
${\bf S_1}$, and ${\bf S_2}$ vectors
are not visible in this figure;
the angles between them can be found
in Fig.~\ref{fig:k_S} in Appendix~\ref{sec:differences}.
}
\end{figure*}
Fig.~\ref{fig:vectors} illustrates the orientation
of electric fields, Poynting vectors, and
wave vector for the two rays leaving the transmitter and
arriving at the receiver for a signal
propagating from the SPICE pulser to A1. For a given ${\bf k}$
of the ray at a given depth,
there are two Poynting
vectors
${\bf S_1}$ and ${\bf S_2}$,
one for each of the two
rays.  These vectors have polar angles
$(\theta_1, \phi_1)$ and $(\theta_2, \phi_2)$,
respectively.
Although the Poynting vectors are often similar
in direction (within a fraction of a degree), 
I keep their directions
different to maintain generality.

For a given ray with Poynting vector 
${\bf S_1}$ or ${\bf S_2}$, in the
absence of cross-polarization power, 
a
$\theta$-type
transmitter emits an
E field polarized
in the ${\bf \hat{{\bm \theta}}^T_1}$ or ${\bf \hat{{\bm \theta}}^T_2}$
direction, respectively.  Each  
${\bf \hat{\bm \theta}}$
direction
is perpendicular to their respective ${\bf S}$
and in the plane of ${\bf S}$  and the
$\gamma$-axis since
the antenna is oriented vertically.

Now we can write an expression
for the total transmitted field 
in terms of the eigenvectors ${\bf E_{1}^T}$
and ${\bf E_{2}^T}$.
In general, 
neither 
${\bf E_1^{T}}$ nor ${\bf E_2^{T}}$
is parallel 
to the ${\bf \hat{\bm \theta}_1^T}$ or
${\bf \hat{\bm \theta}_2^T}$
 directions respectively, but one
will typically be close.  
I call the angle that
${\bf E_{1}^T}$ makes with the ${\bf \hat{\bm \theta}_T}$ direction $ \epsilon_{1}^{T}$.
Then ${\bf E_{2}^T}$
makes an angle $\epsilon_{2}^{T}$ 
with respect to the ${\bf \hat{\bm \phi}}$
direction that is typically
 similar to (within $\lesssim 0.2^{\circ}$ of) $ \epsilon_{1}^{T}$.  For generality, I
maintain both
variables.  These are the same as the $\epsilon$ angles described in Sec.~\ref{sec:epsilons}.
Then the transmitted purely $\theta$-pol
signal is given by the sum of the $\theta$-pol components of the field
from each the two rays:
\begin{align}
    {\bm E^{\rm total}_T} = & {\bm E_1^T} + {\bm E_2^T} \\
    = & 
     E_0 \sin{\theta_{1}^{T}} \cos{\epsilon_{1}^{T}}\,{\bf \hat{E}_{1}^T}  \nonumber \\
     & + E_0 \sin{\theta_{2}^{T}}\sin{\epsilon_{2}^{T}} \,{\bf \hat{E}_{2}^T} .
     \label{eq:txfields}
\end{align}
 In each term in Eq.~\ref{eq:txfields}, the factor containing
$\theta$ is due to the transmitter beam pattern,
and the factors containing $\epsilon$ pick
out the component of the transmitted 
field that is attributed to each 
ray.  Here we are using the beam
patterns of ARA antennas expressed
in Sec.~\ref{sec:assumptions}.  For
ARIANNA, the factors of $\sin{\theta}$
would be replaced with beam patterns
appropriate for LPDAs, but those are
not needed in this paper.

At later times, the power undergoes 1/r and attenuation
losses, and I keep the fraction of power carried by
each ray the same as at the transmitter, but the
directions of 
${\bf E_1}$
and
${\bf E_2}$
change along the rays' path.  The directions
of the E fields change because
the $\epsilon$ angles change with depth, as 
noted in Sec.~\ref{sec:epsilons}.

So, at an 
arbitrary position along the ray's path 
${\bf s}$ and time $t$:
\begin{align}
      r{\bf E^{\rm total}}=& \mathcal{A} \left( {\bf E_{1}} + {\bf E_{2}} \right) \\
      =&\mathcal{A} E_0 \sin{\theta_{1}^{T}}\cos{\epsilon_{1}^{T}}\,{\bf \hat{E}_{1}}  e^{i(\xi_1-\omega t)}  \nonumber \\
    &+ \mathcal{A} E_0 \sin{\theta_{2}^{T}}\sin{\epsilon_{2}^{T}} \,{\bf \hat{E}_{2}} e^{i(\xi_2-\omega t)}
    \label{eq:Efield}
\end{align}
where
\begin{eqnarray}
    \xi_{1,2}=\int_T^{R} {\bf k_{1,2}}\cdot d{\bf s}
\end{eqnarray}
and
\begin{eqnarray}
    r=\int_T^R d {\bf s}, 
\end{eqnarray}
where the limits of integration are the positions
of the transmitter and receiver.
Due to the assumption that
the wave vectors are in the same direction
at all depths, 
 in Eq.(~\ref{eq:Efield}) it is the same ${\bf s}$ that
appears in each term.

\begin{figure*}
    \centering
    \includegraphics[width=1.0\textwidth]{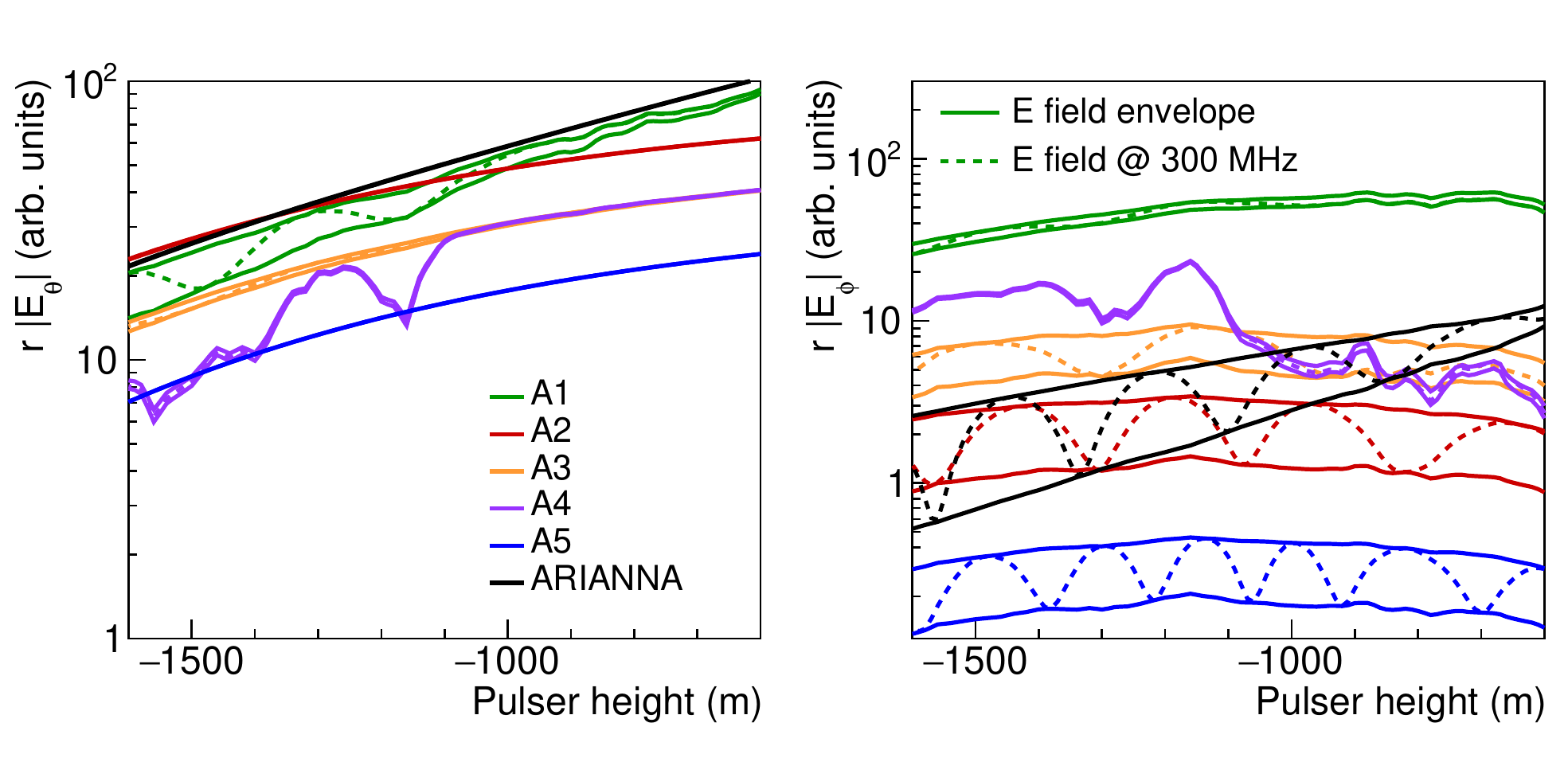}
    \caption{The electric fields expected in
    $\theta$-pol antennas (left) and $\phi$-pol 
    antennas (right) with no power in cross-polarization 
    for each station as a function 
    of pulser height if the signal were purely
    at 300\,MHz.  In an isotropic or uniaxially
    birefringent medium, the signal in 
    $\phi$-pol antennas would vanish.
    The solid lines show the upper and lower bounds
    of the envelope of the field strength, while the dashed
    lines show the field strengths after including the interference term at 300\,MHz.  The same interference
    term subtracts from the power
    in the left plot when it adds to the power in the right plot, and vice versa.  The choice of frequency
    does not affect the envelopes, only the interference terms (dashed lines).  Note the different vertical scales in the two plots.}
    \label{fig:HPolVPolfields}
\end{figure*}
Next, let's consider the time-dependent voltages
seen at a receiver  at a distance.
The (complex) voltage $\mathscr{V}$ at a $\theta$-pol receiver $R$ 
is given by:
\begin{align}
    r\mathscr{V}_{R}=&{\bf E_{1}^R} \cdot {\bf h^R_{\rm eff}}(\theta_{1}^{R},\phi_{1}^{R}) \nonumber \\
    &+{\bf E_{2}^R} \cdot {\bf h^R_{\rm eff}}(\theta_{2}^{R},\phi_{2}^{R})
\end{align}
and using the model for effective
heights described in Sec.~\ref{sec:assumptions},
and ${\bf \hat{\bm \theta}_{1}^R}\cdot {\bf \hat{E}^R_{1}}=\cos{\epsilon_{1}^{R}}$ and 
${\bf \hat{\bm \theta}_{2}^{R}}\cdot {\bf \hat{E}^R_{2}}=\sin{\epsilon_{2}^{R}}$,
the voltage measured at
the $\theta$-type antenna becomes:
\begin{align}
     r\mathscr{ V}_{\rm{\theta}} =&\mathcal{A}E_0 h_0 \left[ \cos{\epsilon_{1}^{T}}\cos{\epsilon_{1}^{R}} \sin{\theta_{1}^{T}} \sin{\theta_{1}^{R}} \,
    e^{i(\xi_1-\omega t)} \right.\nonumber \\
    & \left.+  \sin{\epsilon_{2}^{T}}\sin{\epsilon_{2}^{R}} \sin{\theta_{2}^{T}}
    \sin{\theta_{2}^{R}}\,e^{i(\xi_2-\omega t)} \right] \label{eq:rVtheta}
\end{align}
and the voltage at the $\phi$-type antenna becomes:
\begin{align}
    r{\mathscr{  V}_{\rm{\phi}}} &=\mathcal{A}h_0 E_0 \left[ \cos{\epsilon_{1}^{T}}\sin{\epsilon_{1}^{R}}\sin{\theta_{1}^{T}}\sin{\theta_{1}^{R}} \, e^{i(\xi_1-\omega t)} \right.   \nonumber \\
    &\left.+  \sin{\epsilon_{2}^{T}}\cos{\epsilon_{2}^{R}}\sin{\theta_{2}^{T}}\sin{\theta_{2}^{R}} \, e^{i(\xi_2 -\omega t)} \right].
\end{align}

Now I consider the power at each type of antenna,
using $P_{\theta}=|{\mathscr{ V}_{\theta}}|^2$ and $P_{\phi}=|{\mathscr{ V}_{\phi}}|^2$.
If we define the following
real-valued amplitudes of
the different terms:
\begin{align}
    V_1^{\theta}=\mathcal{A}E_0 h_0 \cos{\epsilon_{1}^{T}}\cos{\epsilon_{1}^{R}} \sin{\theta_{1}^{T}} \sin{\theta_{1}^{R}}  \nonumber \\
    V_2^{\theta}=\mathcal{A}E_0 h_0 \sin{\epsilon_{2}^{T}}\sin{\epsilon_{2}^{R}} \sin{\theta_{2}^{T}}
    \sin{\theta_{2}^{R}} \nonumber \\ 
    V_1^{\phi}=\mathcal{A}h_0 E_0 \cos{\epsilon_{1}^{T}}\sin{\epsilon_{1}^{R}}\sin{\theta_{1}^{T}}\sin{\theta_{1}^{R}}  \nonumber \\
    V_2^{\phi}=\mathcal{A}h_0 E_0 \sin{\epsilon_{2}^{T}}\cos{\epsilon_{2}^{R}}\sin{\theta_{2}^{T}}\sin{\theta_{2}^{R}} 
    \label{eq:realamplitudes}
\end{align}
so that
\begin{align}
    \mathscr{V_{\theta}}=V_1^{\theta} e^{i(\xi_1-\omega t)}+ V_2^{\theta} e^{i(\xi_2-\omega t)} \\
    \mathscr{V_{\phi}}=V_1^{\phi} e^{i(\xi_1-\omega t)}+ V_2^{\phi} e^{i(\xi_2-\omega t)}
\end{align}
then we
can now put the power in each polarization into a simpler form:
\begin{align}
    r^2P_{\theta}=& |r \mathscr{ V}_1^{\theta}|^2 + |r\mathscr{ V}_2^{\theta}|^2 \nonumber \\ 
    +& r^2 V_1^{\theta}V_2^{\theta}\left[e^{i(\xi_1-\xi_2)} + e^{-i(\xi_1-\xi_2)} \right] \\
    =& (rV_1^{\theta})^2 + (rV_2^{\theta})^2  \nonumber \\
    +& 2 r^2 V_1^{\theta}V_2^{\theta} \cos{(\xi_1-\xi_2)} 
    \end{align}
    and then by adding and subtracting $2V_1^{\theta}V_2^{\theta}$, and using
    $\cos{y}=(e^{iy}+e^{-iy})/2$
    and $\sin^2{y}=(1-\cos{2y})/2$, we can write:
    \begin{align}
    r^2 P_{\theta}&=r^2(V^{\theta}_1+V^{\theta}_2)^2\nonumber \\ 
    &- 4r^2V^{\theta}_1 V^{\theta}_2 \sin^2{\left[\dfrac{\xi_1-\xi_2}{2}\right]}.
    \label{eq:power_simple_theta}
\end{align}
Similarly for the $\phi$-type antennas,
\begin{eqnarray}
    r^2P_{\phi}=&r^2(V^{\phi}_1+V^{\phi}_2)^2 \nonumber \\
    &- 4r^2V^{\phi}_1 V^{\phi}_2 \sin^2{\left[\dfrac{\xi_1-\xi_2}{2}\right]}.
    \label{eq:power_simple_phi}
\end{eqnarray}
In Appendix~\ref{sec:terms}, Fig.~\ref{fig:powerplot} shows the
 terms in 
 Eqs.~\ref{eq:power_simple_theta} and~\ref{eq:power_simple_phi}
evaluated for each station observing the SPICE pulses.

Fig.~\ref{fig:HPolVPolfields} shows the electric fields expected at
the location of the $\theta$-pol and $\phi$-pol
receiver antennas as a function of pulser depth for each station at 300\,MHz and with no cross-polarization response in the antennas. 
Fig.~\ref{fig:HPolVPol} in Appendix~\ref{sec:terms} shows
the expected voltages in ARA antennas after the antenna responses
have been folded in.
In the absence 
of a biaxial treatment of birefringence with no cross-polarization
power, the $\phi$-pol antennas would not observe any signal.
We can see that as a function of distance, the most the power
in a $\theta$-pol antenna 
can be is $(V_1^{\theta}+V_2^{\theta})^2$ and
the least it can be is $(V_1^{\theta}-V_2^{\theta})^2$ (and similarly for $\phi$-pol), 
and these bounds form an ``envelope.''  
We can see an interference term 
in the second line of Eqs.~\ref{eq:power_simple_theta} and~\ref{eq:power_simple_phi}, and the
expected voltages including the interference
terms are shown as dashed lines in Fig.~\ref{fig:HPolVPol}.

When the $\epsilon$
angles go to zero and in the 
absence of cross-polarization
power, to a good approximation
the voltages at
the receivers get to the form expected
for the isotropic case.
When
$\epsilon_{1}^{T}=\epsilon_{2}^{T}=\epsilon_{1}^{R}=\epsilon_{2}^{R}=0$, then the voltages
at the two types of receivers become:
\begin{eqnarray}
    r\mathscr{ V}_{\theta}=&&\mathcal{A}E_0h_0\sin{\theta_{1}^{T}}\sin{\theta_{1}^{R}}\,e^{i(\xi_1-\omega t)} \\
    r\mathscr{ V}_{\phi}=&&0,
\end{eqnarray}
and so there is no power in the 
$\phi$-type antenna when we transmit
from a $\theta$-type antenna.  
This 
is the same as the expression we would
find for $r\mathscr{ V}_{\theta}$ for the isotropic
case.  Remember that because of ray bending
due to depth-dependent indices of refraction,
$\theta_{1}^{R} \ne \theta_{1}^{T}$.
I name the following prefactor for the next section:
 \begin{equation}
     \mathcal{F}=\mathcal{A}E_0 h_0 \sin{\theta_{1}^{T}}\sin{\theta_{1}^{R}}.
     \end{equation}

\subsubsection{The Power Envelope and Interference}
\label{sec:interference}
In this section, I discuss important
aspects of the formalism developed 
in Sec.~\ref{sec:propagation}  to develop
intuition.  These are 
the power
envelope and the interference terms.

While power in Eqs.~\ref{eq:power_simple_theta} and
~\ref{eq:power_simple_phi} will oscillate
as a function of distance, they will remain
within an envelope whose bounds are
given by $(V_1^{\theta}+V_2^{\theta})^2$ and
 $(V_1^{\theta}-V_2^{\theta})^2$.  Approximating
 $\theta_{1}^{T}=\theta_{2}^{T}$,
 $\theta_{1}^{R}=\theta_{2}^{R}$, 
 $\epsilon_1^R=\epsilon_2^R=\epsilon^R$,
 and $\epsilon_1^T=\epsilon_2^T=\epsilon^T$,
 the upper bound
 of the envelope becomes:
 \begin{align}
     r^2 P_{\theta}^{\rm env}=&\mathcal{F}^2 \left(\cos^2{\epsilon^T}\cos^2{\epsilon^R}+
     \sin^2{\epsilon^T}\sin^2{\epsilon^R} \right. \nonumber \\
     +&
     \left. 2\cos{\epsilon^T}\sin{\epsilon^T}\cos{\epsilon^R}\sin{\epsilon^R} \right).
     \label{eq:power_theta_simplecase}
 \end{align}
 When both of the
 $\epsilon$ angles are zero,
 this becomes $r^2 P_{\theta}^{\rm env}=\mathcal{F}^2$, as expected for
 the isotropic case.

We have seen that there is also an
interference term.  Using the same
approximations that went into Eq.~\ref{eq:power_theta_simplecase},
 this interference component of
 the power becomes:
 \begin{align}
     r^2P_{\theta}^{\rm inter} =&&-4 \mathcal{F}^2 \cos{\epsilon^{T}}\cos{\epsilon^{R}}
     \sin{\epsilon^{T}}\sin{\epsilon^{R}} \nonumber \\
     && \cdot \sin^2{\left(\dfrac{\xi_1-\xi_2}{2}\right)}.
     \label{eq:interference}
 \end{align}
 Note that this term vanishes if one
 or both of the $\epsilon$ angles is zero.
 
 The presence of the envelope and interference terms
 in Fig.~\ref{fig:HPolVPol} can be understood in terms of the birefringence background that I outlined in Sec.~\ref{sec:background}.
 The variations in the envelope as a function of pulser depth originate from
changes in the $\epsilon$ angles.  The interference term is most important when transmitted signal power is near evenly split between the two eigensolutions.
 
\begin{figure}
\includegraphics[width=0.5\textwidth]{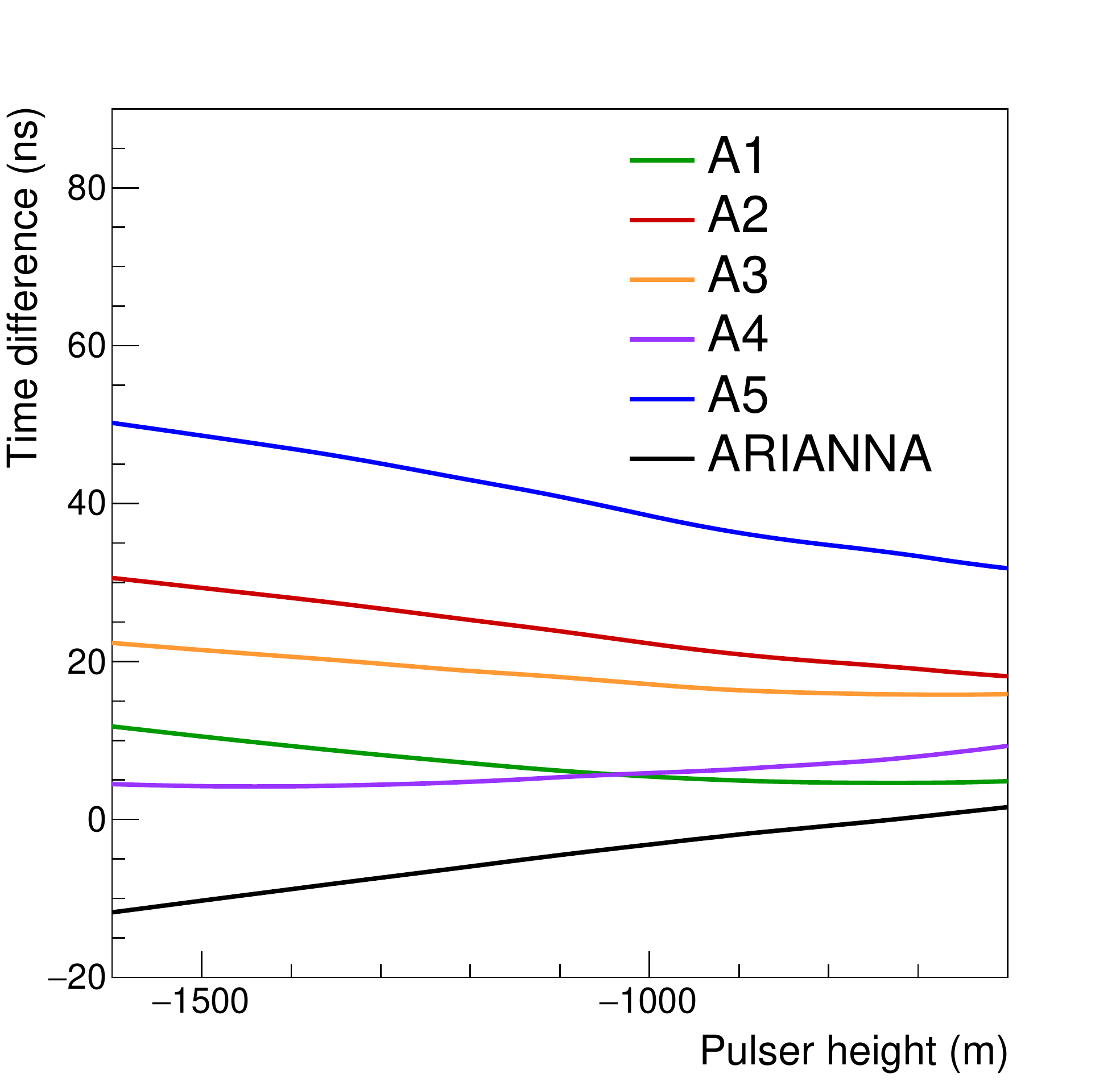}
\caption{\label{fig:timediff} Time difference
between the two rays (both direct rays) 
emitted from the
SPICE transmitter at different pulser
heights upon arrival at each
station.  For the two pulses
to interfere, the width of the pulses 
in time would need to be longer than the 
time differences shown here when incident
on the receivers.  The curve in this plot
representing time differences in A2 is
consistent with what is shown in~\cite{Allison_2020}.}
\end{figure}

 Fig.~\ref{fig:timediff} shows the time difference
 between the signals from the two rays arriving
 at the receivers for the five ARA stations and 
 ARIANNA.
 Note that the two pulses will only interfere
 in time at a receiver if the widths of the arriving pulses are broader than the difference in
 arrival times between them shown in the figure. 

 Even if the pulses do not interfere
 in time, then birefringence will leave an 
 imprint in the received spectrum if both
 pulses are captured in the digitized 
 waveform.
 The argument of the $\sin^2$ function is 
 given by:
 \begin{equation}
     \dfrac{\xi_1-\xi_2}{2}=\int_T^R \left( k_1 -k_2 \right) {\bf \hat{k}} \cdot d{\bf s},
 \end{equation}
where $k_{1,2}=2\pi n_{1,2}f/c$ with $n_{1,2}$ being the indices of refraction
seen by the two rays, which depends on 
${\bf k}$.  So this becomes:
\begin{equation}
     \dfrac{\xi_1-\xi_2}{2}=\dfrac{\pi f}{c}\int_T^R \left( n_{1} -n_{2} \right) {\bf \hat{k}} \cdot d{\bf s}.
 \end{equation}
 If we substitute:
 \begin{equation}
     c\Delta t =  \int_T^R \left( n_{1} -n_{2}\right) {\bf \hat{k}} \cdot d{\bf s}
     \label{eq:pathlength}
 \end{equation}
 where $\Delta t$ is the difference in
 arrival times of the two rays at the 
 receiver,
 then this interference term has nulls when $(\xi_1-\xi_2)/2 =\pi f \Delta t=j \pi$ for integers $j$, or at
 frequencies given by:
 \begin{equation}
     f_{j}=  \dfrac{j}{\Delta t}.
 \end{equation}
  So, for a  distance $d=1$\,km
 and a typical $\Delta n=0.005$, there are nulls in the spectrum every $\Delta f=c/(\Delta n \,d)\approx 60$\,MHz.
 
 If we flip this around, when observable,
 the size of the 
 increments in frequency between nulls is a  measure of the distance to the source if
 the depth-dependent indicatrix is known.
For nulls in a spectrum separated by $\Delta f$,
 the rays have taken a path of length
 given by Eq.~\ref{eq:pathlength}.
 In order for the interference term to be
 observable, the
factor 
$\cos{\epsilon_{1}^{T}}\cos{\epsilon_{1}^{R}}
     \sin{\epsilon_{1}^{T}}\sin{\epsilon_{1}^{R}}$ in Eq.~\ref{eq:interference}, after accounting for cross-polarization,
    needs to be large enough to make
    the deviation from the envelope detectable.  
 
 Ref.~\cite{Matsuoka2009} also
predicts interference patterns, supported by radar data from
Greenland.  They report
significant interference at frequencies of 30\,MHz
and 60\,MHz for radar in 3600\,m and 1200\,m-thick
ice for typical ice fabrics in polar ice sheets.

\subsection{Cross-polarization}

The SPUNK $\theta$-type antenna 
 will also transmit some power in 
 $\phi$-pol, and this is called
 cross-polarization.   Ref.~\cite{Allison_2020} 
 states that from lab measurements,
 the power in cross-pol is reduced to 6\,dB below, or 25\% of, the co-polarization ($\theta$-pol)
 signal power.
 Ref.~\cite{ARIANNA:2020zrg} reports lab measurements of
 cross-polarization signal from the IDL-1
 pulser at 60$^{\circ}$ from the maximum gain of the
 antenna.  From Fig.~3 of that reference, 
 the cross-polarization voltage appears
 to be about 20\% of the co-polarization voltage,
 which would be about 4\% in power.
This fraction of the power in the cross-polarization
may depend on frequency as well as the angle of transmission.

While 
 detectable radio
 emission from in-ice neutrino 
 interactions is expected to be linearly
 polarized with a strong vertical component,
 neutrino-induced emission will not be
 purely vertically polarized.  I expect
 that the same formalism as the one presented here
 can be 
 applied when modeling emission from
 neutrino interactions rather than from 
 pulser antennas.
 
 I simply 
 simulate cross-polarization by
 rotating the polarizations of the transmitted 
 electric field and/or the polarization of 
 the receiver
 by an angle $\delta_{\rm tx}$ ( $\delta_{\rm rx}$) 
 about the direction of ${\bf S}$ for a given ray.  This is a natural
 way to allow for the electric field to have components in two
 perpendicular directions, and is the same
 approach as was used in Ref.~\cite{Adatia1974}.
 The effect on $\epsilon$ angles is additive, 
 such that:
 \begin{equation}
     \epsilon_{1}^{T}\rightarrow \epsilon_{1}^{T}+\delta_{tx},
 \end{equation}
 similarly for $\epsilon_{2,T}$,
 \begin{equation}
     \epsilon_{1}^{R}\rightarrow \epsilon_{1}^{R}+\delta_{rx},
 \end{equation}
 and similarly for $\epsilon_{2}^{R}$.
 For a cross-polarization angle of $\delta=10^{\circ}$,
 the fraction of power in the cross-polarization is $\left(\sin{10^{\circ}}/\cos{10^{\circ}}\right)^2\approx3\%$.
 For $\delta=20^{\circ}$, the fraction is approximately 13\%.

  Cross-polarization can 
  create an effective $\epsilon$ even where one 
  would otherwise be zero.  This leads to 
  interference patterns observable at the
receiver 
that would not have otherwise been present,
as we will see in Section~\ref{sec:data} where
I compare predictions with data.

\section{Comparisons with SPICE Data}
\label{sec:data}
Next, I compare predictions between the model 
described here using published results 
by both the ARA
and ARIANNA experiments from the 
SPICE pulser campaign.  Note that
the framework laid out here is all single-frequency,
and proper comparisons with impulsive events
will require an accounting of the complex
signal spectrum folded in with frequency-dependent effects
such as antenna responses and attenuation in the ice.
The purpose here is to demonstrate that qualitatively,
behaviors are observed in the data that can be brought about 
by signals at
these frequencies propagating in ice
that effectively is biaxially birefringent
 with principal axes used here.

I note a couple of things to keep in
mind for these comparisons.
Where I refer to a model for uniaxial
birefringence, I have set $n_{\beta}$ to the measured $n_{\alpha}$ at a given depth, with an axis of symmetry about the ${\gamma}$-axis.  Also, I include the interference term, which is valid for modeling spectra as long as signals from both rays are contained in a measured waveform.  For quantities measured in the time
domain such as peak voltage, the effect of the interference term will be more complicated if the pulses are only partially overlapping or not overlapping at all.
As can be seen in Ref.~\cite{Allison_2020}, the width of the SPICE pulses observed in ARA is approximately 50\,ns.

\subsection{Non-trivial rotation of polarizations}
\begin{figure*}
    \centering
    \includegraphics[width=1.0\textwidth]{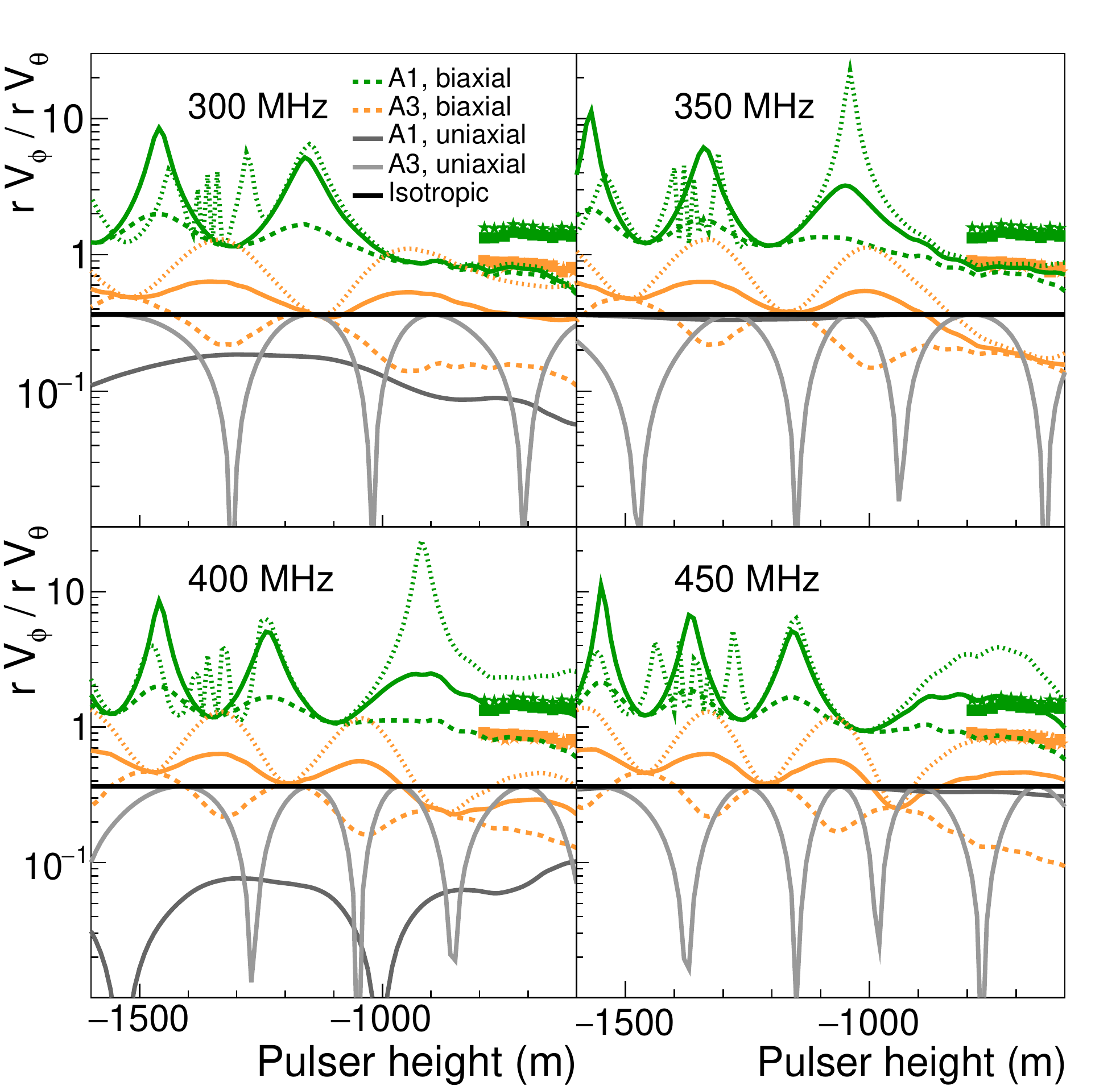}
    \caption{Ratios of signal voltages expected in $\phi$-pol
    and $\theta$-pol at single frequencies 
    in stations A1 and A3 compared with measured ratios of voltage SNRs in the two polarizations (points).
    The lines are for antenna cross-polarization
    angles
    $\delta_{\rm rx}=-\delta_{\rm tx}=0^{\circ}$ (dashed), $10^{\circ}$ (solid), and
    and $20^{\circ}$ (dotted).  Note that I chose solid lines for $\delta_{\rm rx}=-\delta_{\rm tx}=10^{\circ}$.
    The square markers are from the top pairs of $\theta$-pol, and $\phi$-pol antennas in a station, and the stars from the bottom pairs.
    The takeaway is that the high 
    $rV_{\phi}/rV_{\theta}$ ratios observed from SPICE pulses in ARA stations
    are difficult to achieve with either uniaxial
    birefringence or isotropic ice, but are
    achievable with a biaxial treatment of birefringence
    at some frequencies for some choices of
    $\delta_{\rm rx}$ and $\delta_{\rm tx}$. The $\delta$ angles are the only free parameters in these predictions. 
    \label{fig:ara_voltage_ratios} }
\end{figure*}

In Fig.~\ref{fig:ara_voltage_ratios}, I
compare the predicted 
ratio of voltages in the two polarizations
($V_{\phi}/V_{\theta}$) with ratios
of SNRs derived from Figs.~10 and 12 in Ref.~\cite{Allison_2020} for two ARA stations,
A1 and A3 (although ratios of voltages and ratios of
SNRs are not the same, for SNRs of about 10 as in Figs.~10 and 12, comparisons of 
SNRs and amplitudes
are similar at the ${\sim}10$\% level). 
We compare at four frequencies: 300\,MHz, 350\,MHz, 400\,MHz, and 450\,MHz.  Ref.~\cite{Allison_2020} reports
measured signal SNRs in the four 
top and bottom
$\theta$-pol and $\phi$-pol antennas,
respectively. 
I compare with data points from
the greatest pulser depths in the plot up 
to the shadow zone boundary at 600\,m
depth so that we are only comparing
direct signals.  
Fig.~\ref{fig:ara_voltage_ratios} includes
predictions
for different choices
of antenna cross-polarization angles, with the solid lines representing $\delta_{tx}=10^{\circ}$ and 
$\delta_{rx}=-10^{\circ}$.  
Note that these $\delta$ angles
are the only free parameters in this model.
We see that the model for pulses
transmitted and received in ice
that is effectively biaxially
birefringent leads to $\phi$-pol power exceeding $\theta$-pol power, as observed in
the data, at all four frequencies in A1.  
The reason for this is that for 
SPICE signals that reach A1,
the 
epsilon angles that the signal sees at the receiver differ greatly from those that it sees
at the transmitter, causing the
signal polarization to rotate
correspondingly.

Fig.~\ref{fig:ara_voltage_ratios} also includes
uniaxial and isotropic 
models.  In a uniaxial crystal, the
$\epsilon$ angles vanish and do not change along
the rays' path, and so the power envelopes
do not change, but we still expect interference
between the two rays.  So, the structure observed
in the uniaxial models is from interference alone.  In an isotropic medium, we expect
that the power observed in each polarization 
at the receiver is
the same as they were
at the transmitter, which for 
the choice of $\delta_{tx}=-\delta_{rx}=10^{\circ}$
is $(\sin{10^{\circ}}/\cos{10^{\circ}})^2~\approx~3$\%.

Although I do not claim that  Fig.~\ref{fig:ara_voltage_ratios} shows
 good agreement between the model and the
data at any single frequency for both stations, qualitatively it looks promising.
The model does predict $V_{\phi}/V_{\theta}>1$ in A1,
as is observed
in the data, for
some frequencies in ARA's band for reasonable cross-polarization angles 
$\delta_{rx}$ and $\delta_{tx}$.  In A3, $V_{\phi}/V_{\theta}$
ratios approaching unity are achieved for
some frequencies as well, but for higher
 cross-polarization fractions corresponding
 to $\delta$ angles of around 20$^{\circ}$.
With uniaxial birefringence, 
this fraction does not
come as close to the data points,
and the data and prediction
are even more discrepant for
isotropic ice.

\subsection{Interference}

\begin{figure}
    \centering
    \includegraphics[width=0.5\textwidth]{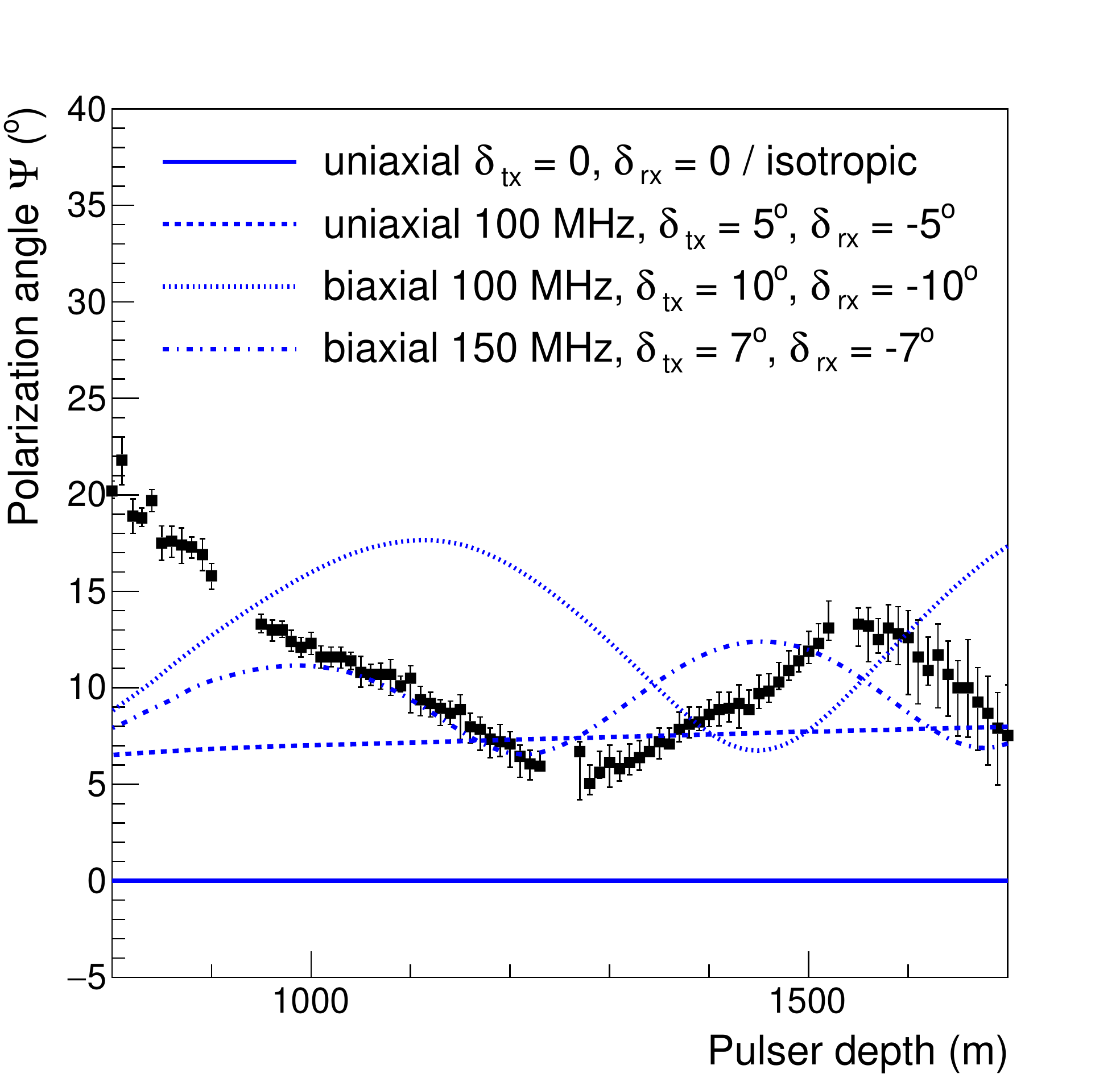}
    \caption{Comparison of measured polarization angle $\Psi$ of the SPICE pulser in the ARIANNA South Pole station with predictions at single frequencies in ARIANNA's band and for different choices of the percentage of power in cross-polarization.  I also show the case of uniaxial birefringence for $\delta_{tx}=-\delta_{rx}=5^{\circ}$ and for no cross-polarization.  For pulser depths
more shallow than 938\,m there is expected to be interference
between direct and reflected rays~\cite{Geoff2021}, so comparisons are only valid
for pulser depths $>938$\,m.}
    \label{fig:polarization_arianna}
\end{figure}

In Fig.~\ref{fig:polarization_arianna},
I compare predictions to the 
polarization angle $\Psi$ reported by 
ARIANNA in Ref.~\cite{ARIANNA:2020zrg} as a function of
pulser depth.  The polarization angle in Ref.~\cite{ARIANNA:2020zrg}  is related to
the ratio
of voltages in two polarizations by:
\begin{equation}
\Psi=\tan^{-1}{\left(|\mathscr{V}_{\phi}|/|\mathscr{V}_{\theta}|\right)}.
\end{equation}
For this measurement, only the LPDA
antennas measuring polarization in the horizontal
plane were used~\cite{Geoff2021} and the 
paper refers to $\mathscr{V}_{\phi}$ and $\mathscr{V}_{\theta}$
as being from the fields perpendicular to the direction of propagation.
Again, the only free parameters in
the model predictions in this plot
are the choices of $\delta_{\rm tx}$
and $\delta_{\rm rx}$.
Ref.~\cite{Geoff2021} states that for pulser depths
more shallow than 938\,m there is expected to be interference
between direct and reflected rays, so comparisons are only valid
for pulser depths $>938$\,m.
As can be seen in the figure,
the prediction is frequency-dependent, but the observations
are of the size expected for cross-polarization angles $\delta_{\rm rx}$ and $\delta_{\rm tx}$ of several
degrees.

Also shown
is the expectation for a uniaxial ice crystal
with cross-polarization included,
which does not show the dramatic depth-dependent structure.  The way I set the parameters of
the uniaxial crystal for these comparisons
gives a long distance for the oscillations, making the
uniaxial curve on this plot nearly flat.
For the uniaxial
case with no cross-polarization, the $\Psi$
angle vanishes, also shown.  
Note the vertical axis goes 
negative to make this visible.

In Ref.~\cite{ARIANNA:2020zrg} ARIANNA 
 also shows a measured SPICE
pulser spectrum that does not appear 
to agree in shape with the one measured
in the laboratory, with dips 
every ${\sim}75$\,MHz (see their Fig.~7).  Recall that 
dips in the spectrum are predicted
in increments of $\Delta f=1/\Delta t$.
From the time differences
in Fig.~\ref{fig:timediff}, at the greatest
pulser depths I would
expect oscillations in the spectrum
 every ${\sim}1/(\Delta t)=1/12$\,ns=83\,MHz. While this is suggestive, a more
 complete model of the station
 so near the surface may be needed to compare the measured spectrum with an expectation derived from the model in this paper.

\begin{figure}
\includegraphics[width=0.5\textwidth]{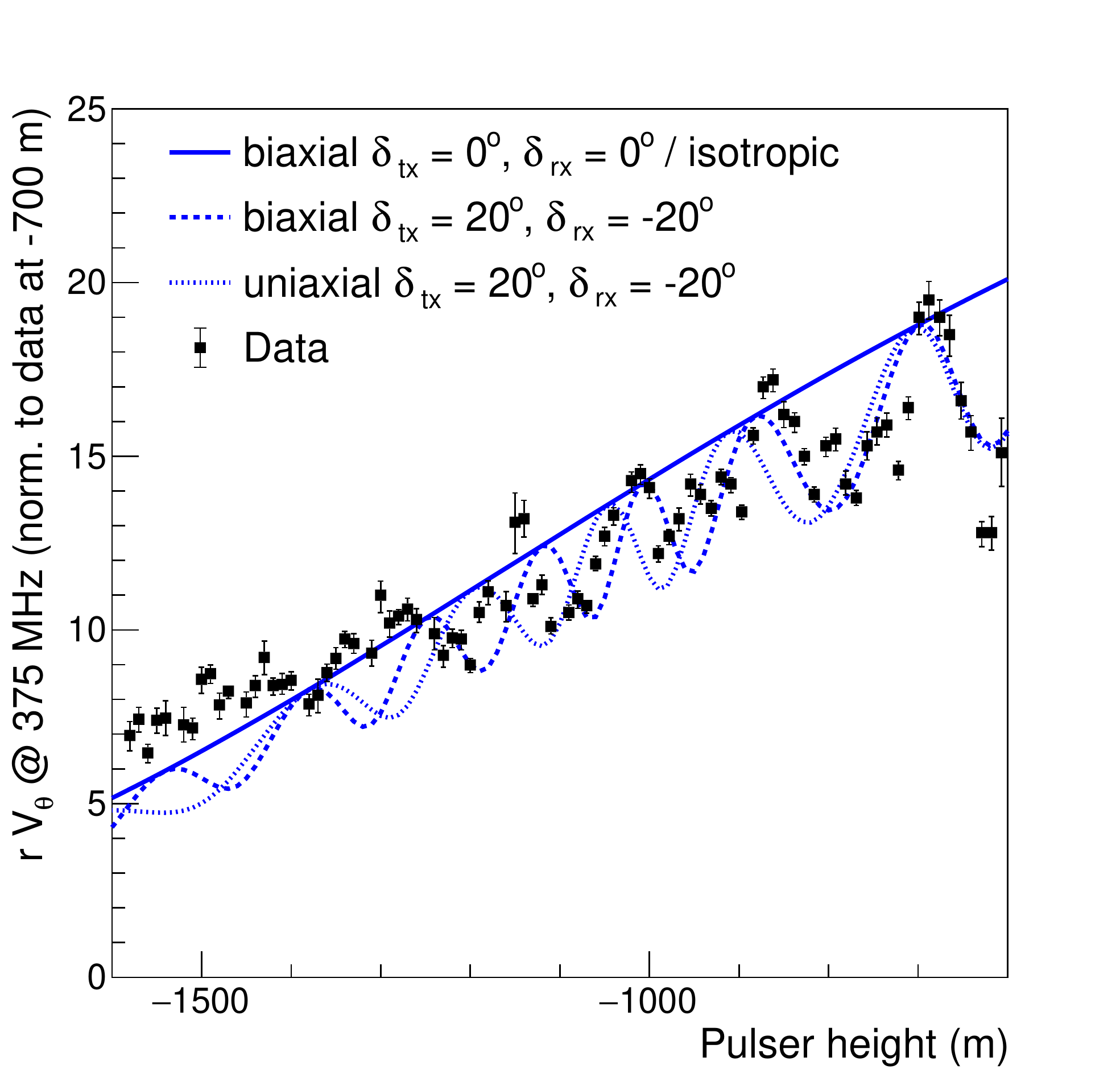}
\caption{\label{fig:voltages_a5} Peak voltage
measured in
 $\theta$-pol antennas 
at the A5 station as a function of SPICE pulser height, compared to predictions
at 300\,MHz for different choices
of cross-polarization power fraction.  The
models are scaled to the measurement
at 700\,m pulser depth.
}
\end{figure}

In Fig.~\ref{fig:voltages_a5}, I compare
the predicted voltages in $\theta$-pol
antennas in A5 with those reported in 
Fig.~13 of Ref.~\cite{Allison_2020}.
The model shows oscillations of the
magnitude seen in the data whether
the model uses a uniaxially or biaxially birefringent crystal and for
cross-polarization angles of
about $\delta_{\rm tx}=-\delta_{\rm rx}=20^{\circ}$.
These oscillations are due to 
interference between the rays 
at the receiver and rely on the pulses
from the two rays overlapping in time
on arrival at the receiver.  Due to the small $\epsilon$ angles for station A5 seen in Fig.~\ref{fig:epsilons_sidebyside},
variations in the signal envelope
are small, and variations are dominated instead by interference.
In this case, the 
interference is present whether the crystal is
modeled as uniaxially or biaxially birefringent.

Additionally, I note that in Ref.~\cite{Barrella:2010vs}, similar
periodic variations in the power spectrum
every ${\sim}200$\,MHz
are also reported in signals that were transmitted
from the surface of the Ross Ice Shelf
and received after reflections from
the ice-sea water interface below.
They also report observing power in
the cross-polarization that is
frequency-dependent and peaks
at 80\% of the total received power
at 450\,MHz, and that was only seen
in in-ice measurements and not those taken in
air.  The authors do suggest that the 
modulation could be due to birefringence,
with two signals arriving at different
times due to wave speeds that differ by 0.1\%.
For the measurements reported
in Ref.~\cite{Barrella:2010vs}, signals were transmitted vertically,
and under a model where the vertical
axis is a principal axis of the indicatrix,
there would still be two extraordinary
rays that could interfere and cause
the structure in the spectrum in the
co-polarization.
However, the $\epsilon$ angles would vanish, and so under that model, power in the
cross-polarization would not appear
beyond what was transmitted.
Still, Ref.~\cite{Matsuoka2009} notes
that on ice shelves, in deep ice, and ice
caps near the coast, the
crystal fabric can have a more complicated
structure due to being warmer and 
strains being more complex.
Ref.~\cite{Barrella:2010vs} recommended
that 
future measurements be taken at many
angles with respect to the ice fabric.

Lastly, I note that in Ref.~\cite{Allison:2011wk}, ARA reported
calibration pulses nominally 
transmitted in $\theta$-pol and 
observed after propagating approximately 3.2\,km
in the ice with a significant $\phi$-pol component,
about a factor of 2-3 weaker in amplitude.  The higher-than-expected
$\phi$-pol power was attributed to
radiation of cross-polarization power
at the transmitter due to challenges
in antenna construction during
rapid deployment.  Birefringence
could be investigated as contributing
to the observed $\phi$-pol power.

In summary, the model presented
in this paper can bring about
expected behaviors that are
difficult to explain without a 
biaxial treatment of birefringence.  However,
we have seen that the same model
parameters do not give a best
fit to all of
the data at once.  For example,
Fig.~\ref{fig:ara_voltage_ratios}
and Fig.~\ref{fig:polarization_arianna}
seem to prefer  cross-polarization angles $\delta_{rx}$
and $\delta_{rx}$ of approximately $10^{\circ}$, while
Fig.~\ref{fig:voltages_a5}
seems to indicate a need for 
higher cross-polarization angles.  This could be due to
one or more aspects of the model
being incomplete.
In addition to the aforementioned
need for a broadband treatment, the indicatrix may be oversimplified,
as is surely the method 
used to model
cross-polarization.

\section{Future Work}
Further work is needed for more 
quantitative comparisons with the 
broadband signals in the SPICE dataset.
For example, true antenna beam patterns
of all types as well as the true complex 
pulse spectrum should be included in
the model.  A ray-tracing
algorithm for propagating both direct and
refracted signals
in biaxially birefringent ice is also 
needed. Ideally, a finite-difference
time-domain simulation could be used to 
validate a more complete model, although
it would need to be tested over shorter distances 
(100s of meters)
for manageable computational times~\cite{RadarEchoTelescope:2020nhe}. 
In addition, the model needs to
be made more general to loosen the assumption
that two axes of the indicatrix are perfectly
aligned with the vertical direction and along
the direction of ice flow, since 
measurements from the ice report that they
could be different by approximately 10$^{\circ}$~\cite{Matsuoka2009}.

\section{Conclusions}
Data from the SPICE pulser program
strongly suggest that a biaxial treatment
of 
birefringence in the ice near South Pole,
with principal axes defined by the directions
of ice flow and vertical compression,
has an important effect on the power
of signals observed by antennas 
in the ice measuring different polarizations.
Observed power in the two polarizations 
has a non-trivial
dependence on the
positions of the transmitter and receiver 
and the orientation of the pair relative
to the principal axes of the birefringent
crystal.  While deviations compared
to the isotropic or uniaxial expectations would be
present even if the transmitter were to emit
purely in one polarization, they are
enhanced when there is some cross-polarization power transmitted and received.

While the effects described here add
complexity to  neutrino detection using
radio techniques in in-ice detectors,
the same effects provide important
signatures for the identification of signals
originating from neutrinos in the ice.
For example, the arrival of two rays
with the expected time differences of order
tens of ns from nearly the same direction
will be an important signature of
an in-ice interaction.
Then, the distance 
to the source can be traced using
the difference in time arrival between
the two rays (evident in either the
time or frequency domains).  This will
broaden the category of events whose 
distance can be reconstructed more precisely than those relying on wavefront curvature, alongside
the ``double-pulse'' events that contain
both direct and refracted pulses.  The
distance to the interaction is
crucial for reconstructing the energy
of the neutrino-induced shower.
Birefringence that is effectively
biaxial at radio frequencies also needs to 
be handled properly to
reconstruct the 
polarization of the signal at the interaction,
which is required for reconstructing the direction of the incident neutrino source.

Neutrino directional reconstruction
will require some knowledge of
 the  birefringence parameters
along the signal's path.
These parameters could be derived in
advance
from in-ice pulser data like SPICE (for in-ice experiments),
or from radar data such as that
from CReSIS~\cite{cresis}. Refs.~\cite{tc-15-4117-2021,fujita_maeno_matsuoka_2006,Matsuoka2012,Young2021,Brisbourne2019,Jordan2019,Dall2010,jordan_schroeder_elsworth_siegfried_2020}
have demonstrated that
radar polarimetry can be used
to extract properties of the COF 
as well as from measurements taken through the more
laborious extracting of ice cores.
Seismology measurements have also been
used to extract properties of the
COF~\cite{horgan_anandakrishnan_alley_burkett_peters_2011,Brisbourne2019,velez}.  Alternatively,
perhaps enough information about
the crystal could be extracted from
the neutrino signal itself to enable
its reconstruction.

The effects described here will also 
impact the design of detectors.
 Some care will need to be taken
to optimize an array for the greatest science
potential, which may be a balance 
between the desire to detect high power in one polarization (for example, in-ice detectors
can more easily measure $\theta$-pol) 
and the desire to observe 
signatures that are more visible in 
the other polarization (for example, oscillations are more apparent in $\phi$-pol).  The most optimal 
arrangement of antennas in an in-ice array
may or may not have a symmetry around
the direction of ice flow.  
For example, since changing polarizations due to
rotating eigenstates is maximal in the $\alpha$-$\gamma$ plane, detectors could be designed
to view away from that plane
to avoid such large variations.  On the other hand,
if such variations could be exploited to improve
reconstruction, then detection of interactions from the $\alpha$-$\gamma$  plane would be desirable.
Similarly, the importance of viewing the time differences
between the two rays, which at South Pole are maximal for directions along the $\beta$-$\gamma$ plane and can be used for distance reconstruction, should 
also be considered.
While
the $\phi$-pol antennas deployed by 
ARA are adequate to have observed the
effects here, in future arrays  the 
sensitivity to the
$\phi$-pol component of the signal 
should not be
diminished without careful consideration.

\section{Acknowledgements}
  Thank you to the SPICE
team as well as the members of the ARA and
ARIANNA collaborations for their contributions
in producing the SPICE
pulser data set, which is an excellent one
for exploring these effects.
I am grateful to Dave Besson
for many helpful discussions on this topic and his review of the draft, and Steven Prohira 
for useful feedback as my thinking has evolved
on how to model these effects and for important feedback on improving the clarity of the paper.  Thanks to Jorge
Torres for his work on polarization reconstructions
that led me to think about these behaviors in
the data.
Thank you to S. Bektas for promptly responding with
the word document of his paper for finding
the ellipsoid intersection.
Thank you to Justin
Flaherty for helping me with some of the fundamentals of birefringence, and
for feedback on the paper draft.
Thank you as well to John Beacom,
Dima Chirkin, William Luszczak,
Chris Hirata, Roland Kawakami, 
and Martin Rongen for reviewing the draft and providing valuable feedback.
Thanks to James Beatty, Nicholas
Harty, Dave Seckel, Ilya Kravchenko, Patrick Allison,
Peter Gorham, and other members
of the ARA and PUEO Collaborations as well 
for discussions and feedback.
Thanks additionally to Kenneth Jezek, Prasad Gogineni, and Carlos Martin Garcia.
 Any inaccuracies or
errors are my own.
This work was supported by National Science Foundation award 1806923.

\appendix

\section{Special directions in South Pole ice}
\label{sec:V}
In a biaxially birefringent crystal,
there are only two special directions of ${\bf k}$
where waves behave the same as in an isotropic medium,
with only a single ray with one index of refraction,
${\bf S}\parallel {\bf k}$, and ${\bf E}\parallel {\bf D}$.
These two special directions of ${\bf k}$  sit in the $\alpha-\gamma$
plane and are the two directions 
for which the intersection of the planar
wavefront with the indicatrix
becomes a circle.  These two special directions make an angle $V_{z}$ with respect to the $\gamma$-axis
given by~\cite{StoiberMorse}:
\begin{equation}
    \cos{2V_z} = \dfrac{(n_\gamma-n_\beta) - (n_\beta-n_\alpha)}{(n_\gamma-n_\alpha)}.
\end{equation}
\begin{figure}
\includegraphics[width=0.5\textwidth]{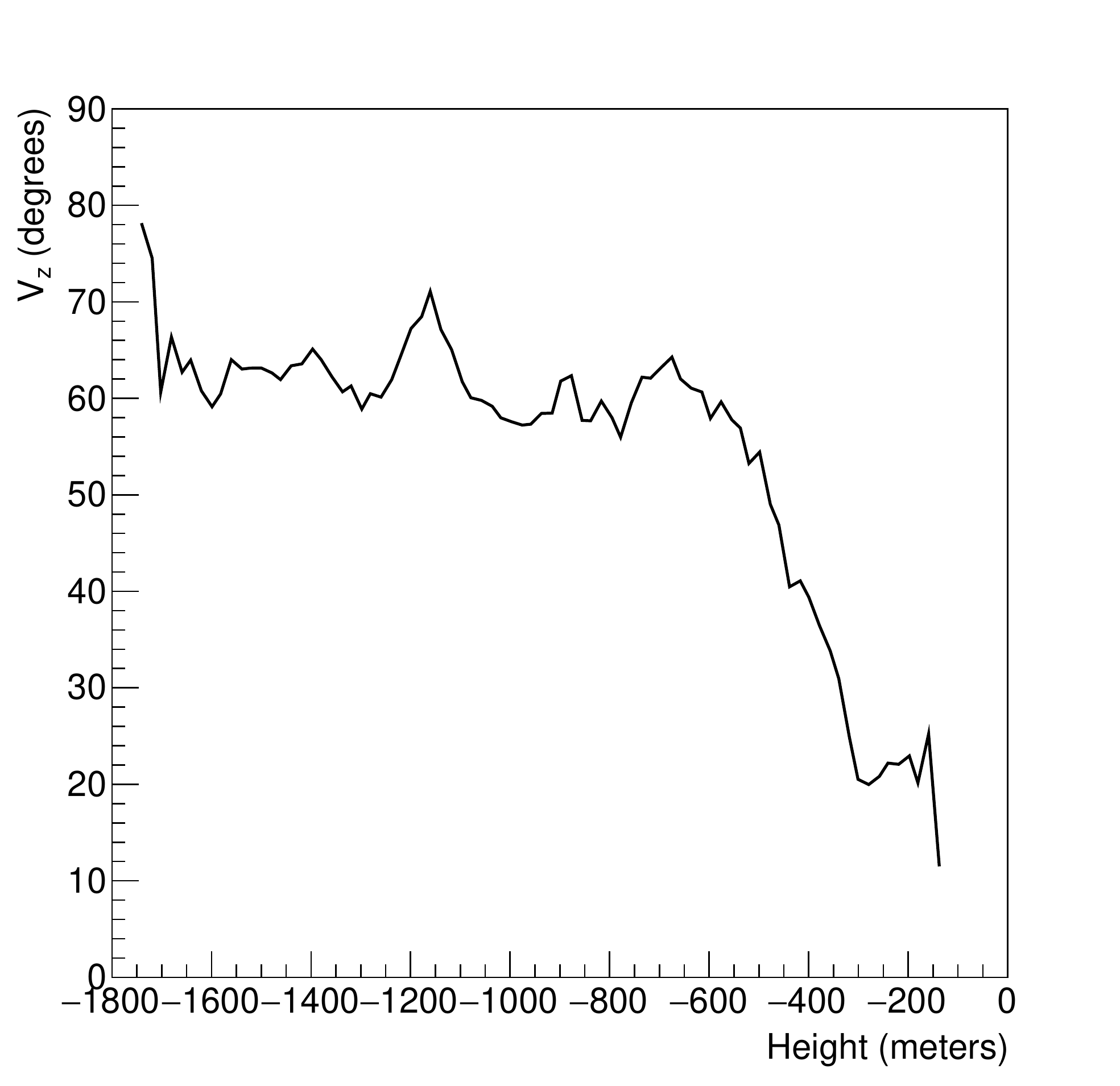}
\caption{\label{fig:V} Treating
the ice as effectively a biaxially
birefringent medium at radio
frequencies, there are two special
directions where rays propagate as they 
would in an isotropic medium.  These
directions are in the plane of the
ray and the 3-axis, each making an angle
 $2V_z$ with the 3-axis.  The angle $V_z$
 is shown here as a function of height
 in South Pole ice.}
\end{figure}

Fig.~\ref{fig:V} shows the $V_z$ angle
as a function of depth in the ice using the indicatrix parameters shown in Fig.~\ref{fig:n123}.  I include this plot
as an interesting
piece of information about South Pole ice.

\section{Power and voltages}
\label{sec:terms}
\begin{figure*}
    \centering
    \includegraphics[width=1.0\textwidth]{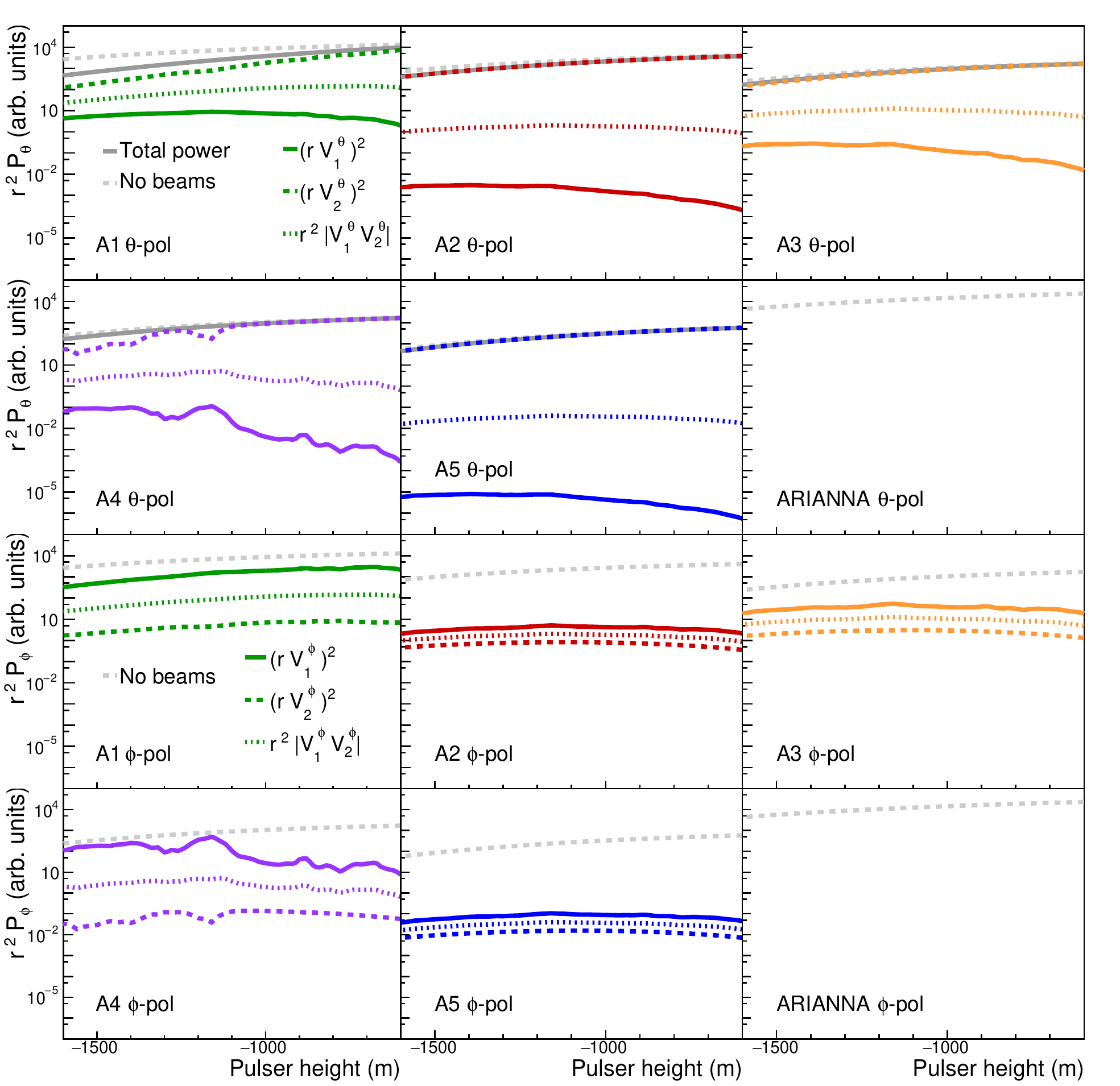}
    \caption{ These figures show the terms
    contributing to the power at each 
    station predicted to be 
    measured in $\theta$-pol (top eight panels) and $\phi$-pol
    antennas (bottom eight).  These were made
    at 300\,MHz with no cross-polarization power in the antennas, just as in Fig.~\ref{fig:HPolVPol}.}
    \label{fig:powerplot}
\end{figure*}

\begin{figure*}
    \centering
    \includegraphics[width=1.0\textwidth]{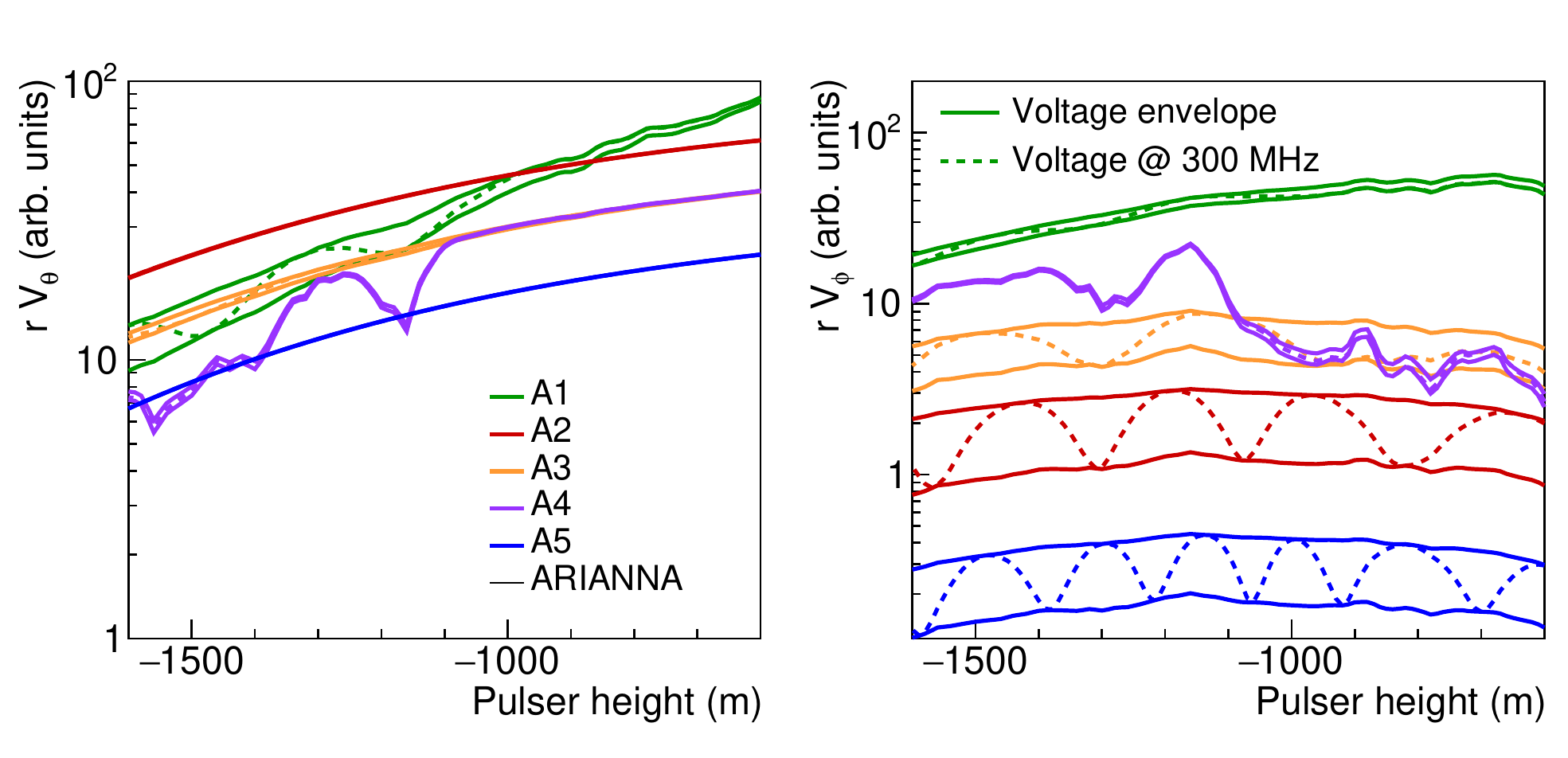}
    \caption{The voltage amplitudes expected in
    $\theta$-pol antennas (left) and $\phi$-pol 
    antennas (right) with no power in cross-polarization 
    for each station as a function 
    of pulser height if the signal were purely
    at 300\,MHz.  In an isotropic or uniaxially
    birefringent medium, the signal in 
    $\phi$-pol antennas would vanish.
    On the left, the lines show the upper bounds of the voltage envelope.
    On the right,
    the solid lines show the upper and lower bounds
    of the voltage envelope, while the dashed
    lines show the voltages after including the interference term at 300\,MHz.  The same interference
    term subtracts from the power
    in the left plot when it adds to the power in the right plot, and vice versa.  Note the different vertical scales in the two plots.}
    \label{fig:HPolVPol}
\end{figure*}

In Fig.~\ref{fig:powerplot}, I include
the terms contributing to the power in
$\theta$-pol (top eight plots) and $\phi$-pol (bottom eight) in Eqs.~\ref{eq:power_simple_theta} and~\ref{eq:power_simple_phi}
for each ARA station.  This plot was made
at 300\,MHz frequency with $\delta_{\rm tx}=\delta_{\rm rx}=0$.  
For $\theta$-pol, solid gray lines represent the total power and  dashed gray lines represent the
 same but with antenna responses removed.
 The analogous quantities exist in
$\phi$-pol as well but are not shown.
Fig.~\ref{fig:HPolVPol} shows the voltages expected in
the ARA receivers calculated from the electric fields
in Fig.~\ref{fig:HPolVPolfields} and folding in the 
antenna responses in Eq.~\ref{eq:effectiveheight}.

\section{Comparing quantities in the two eigensolutions}

\label{sec:differences}
In this section, I compare similar quantities in the two
different eigenstates for a given
$\vec{k}$ at a given depth relevant
to the SPICE pulser data.  In this paper,
both rays representing the two
eigensolutions are given the 
same ${\bf k}$ at each depth, but
the direction of the corresponding
Poynting vectors ${\bf S_1}$ and
${\bf S_2}$ differ slightly, as
do the orientation of the electric fields
about the Poynting vectors, 
$\epsilon_1$ and $\epsilon_2$.

\begin{figure}
    \centering
    \includegraphics[width=0.49\textwidth]{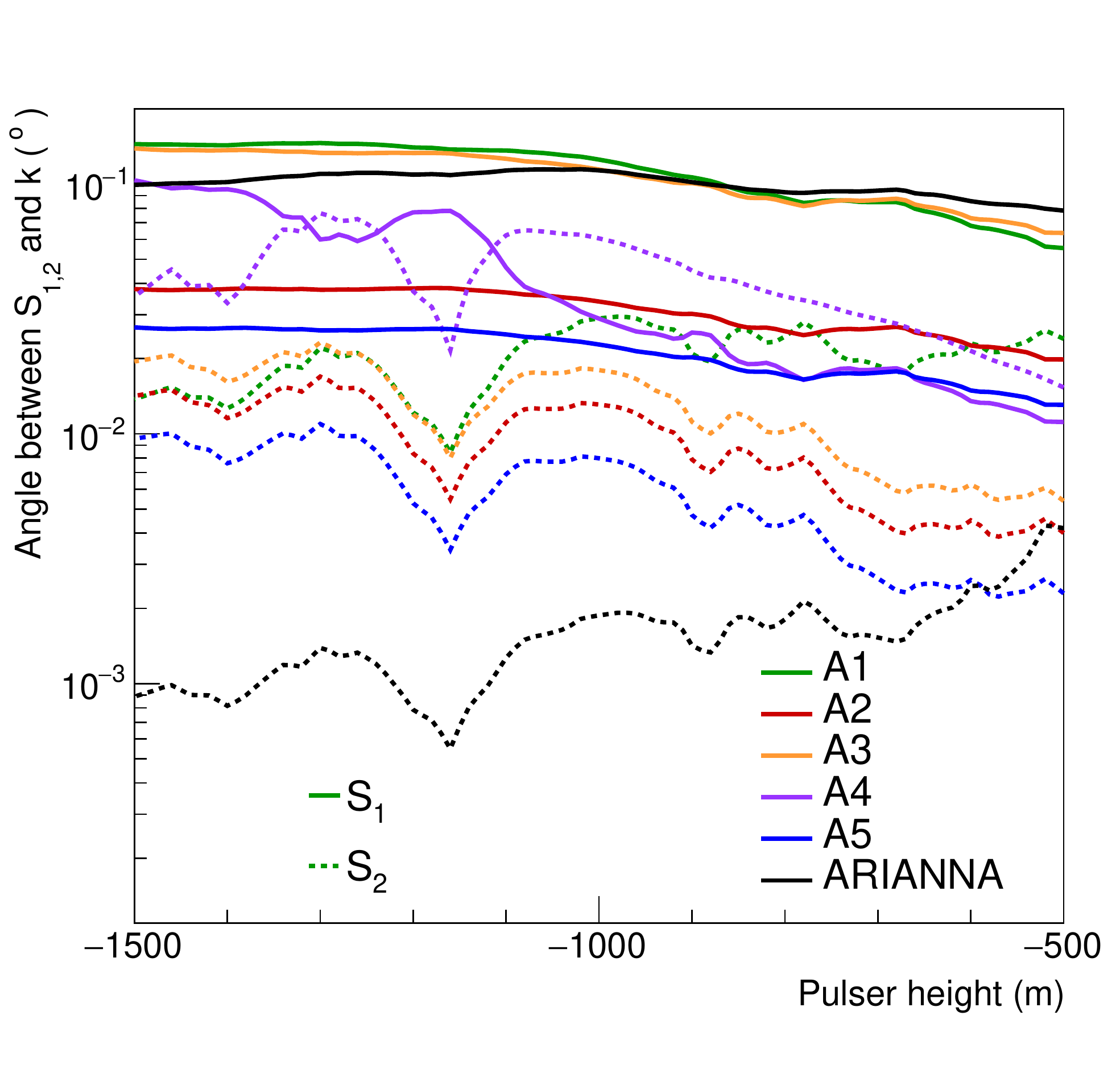}
    \caption{Angle between the wave vector $\vec{k}$
    and the Poynting vectors  $\vec{S}_{1,2}$ for the two
    corresponding eigenstates at the transmitter, as a function of depth of the pulser.}
    \label{fig:k_S}
\end{figure}

Fig.~\ref{fig:k_S} shows the angles
between ${\bf k}$ and each of ${\bf S_1}$ and ${\bf S_2}$ at the transmitter as a function of pulser height.  We can see that this angle 
is sub-degree for all pulser depths.  In this paper, I take the angles relevant
to the antenna responses to be in the
direction of the Poynting vectors rather
than ${\bf k}$, and the direction
of the propagation of the rays to
be in the ${\bf k}$ direction, not ${\bf S}$.  This figure shows that these
choices will have a negligible impact
on the results in this paper and
although would be interesting to validate
in the future with data, would be challenging given angular resolutions
of experiments.

\begin{figure*}
    \centering
    \includegraphics[width=0.98\textwidth]{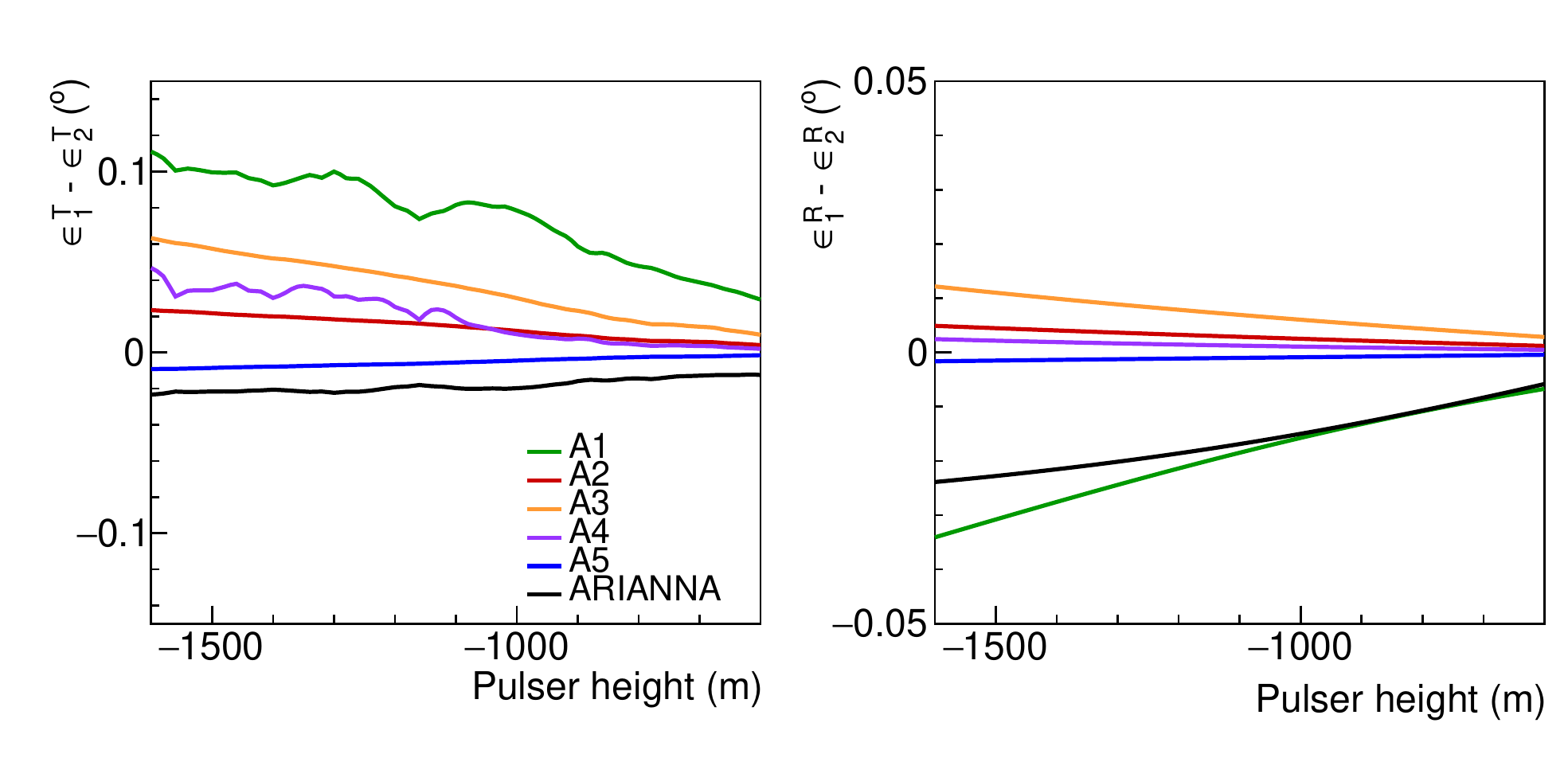}
    \caption{Angle between the $\epsilon$ angles for the two eigensolutions at the transmitter (left) and receiver (right).}
    \label{fig:diffepsilons}
\end{figure*}

Fig.~\ref{fig:diffepsilons} shows the
difference between epsilon angles for the two eigenstates at the transmitter and 
receiver.  These angles are also similar
to one another to within less than
about $0.1^{\circ}$.

\label{sec:k_S}

\section{Dependence of $\epsilon$ angles on direction of ${\bf k}$}

\begin{figure*}
    \centering
    \includegraphics[width=0.48\textwidth]{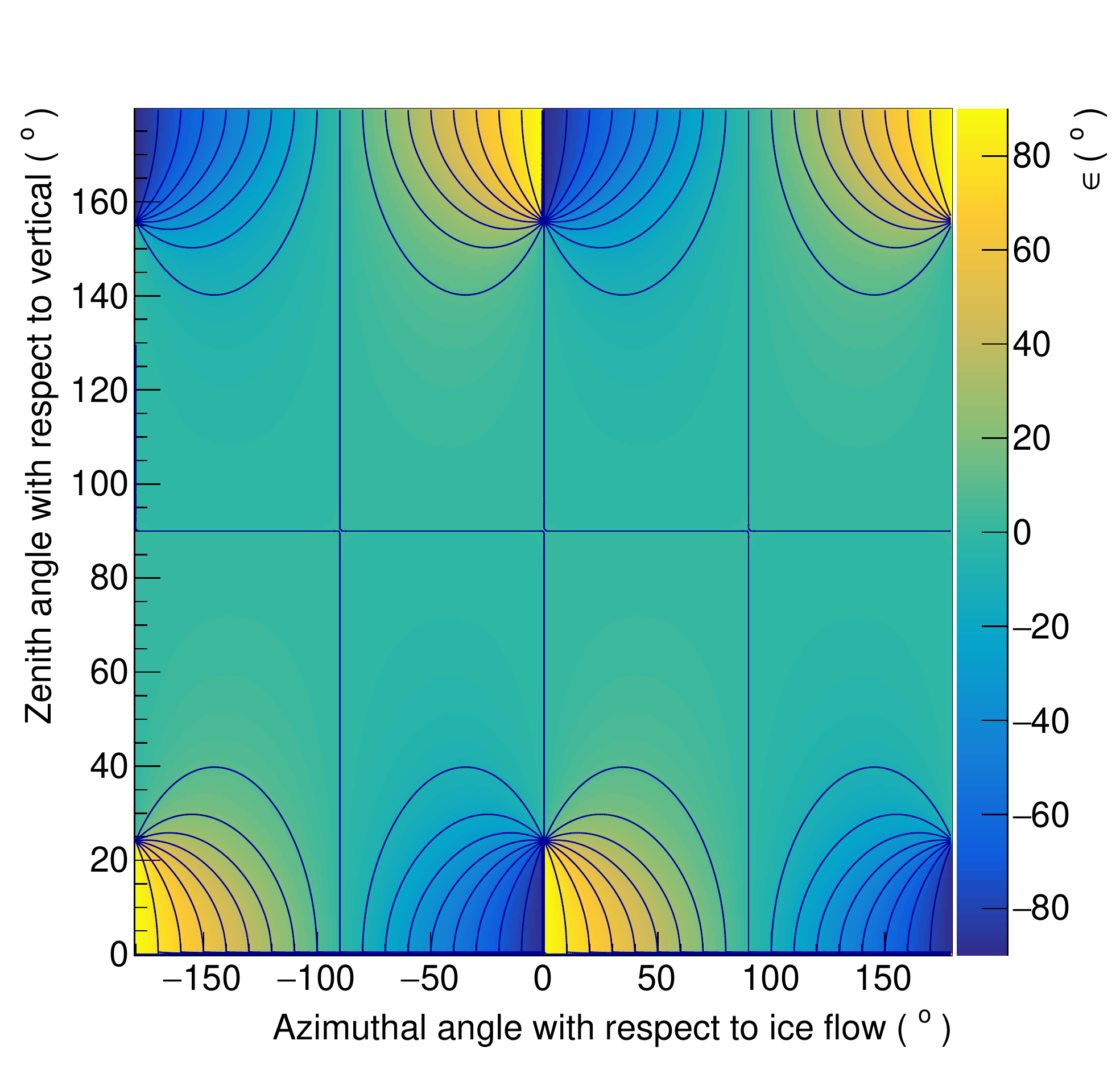}
    \hspace{0.2in}
    \includegraphics[width=0.48\textwidth]{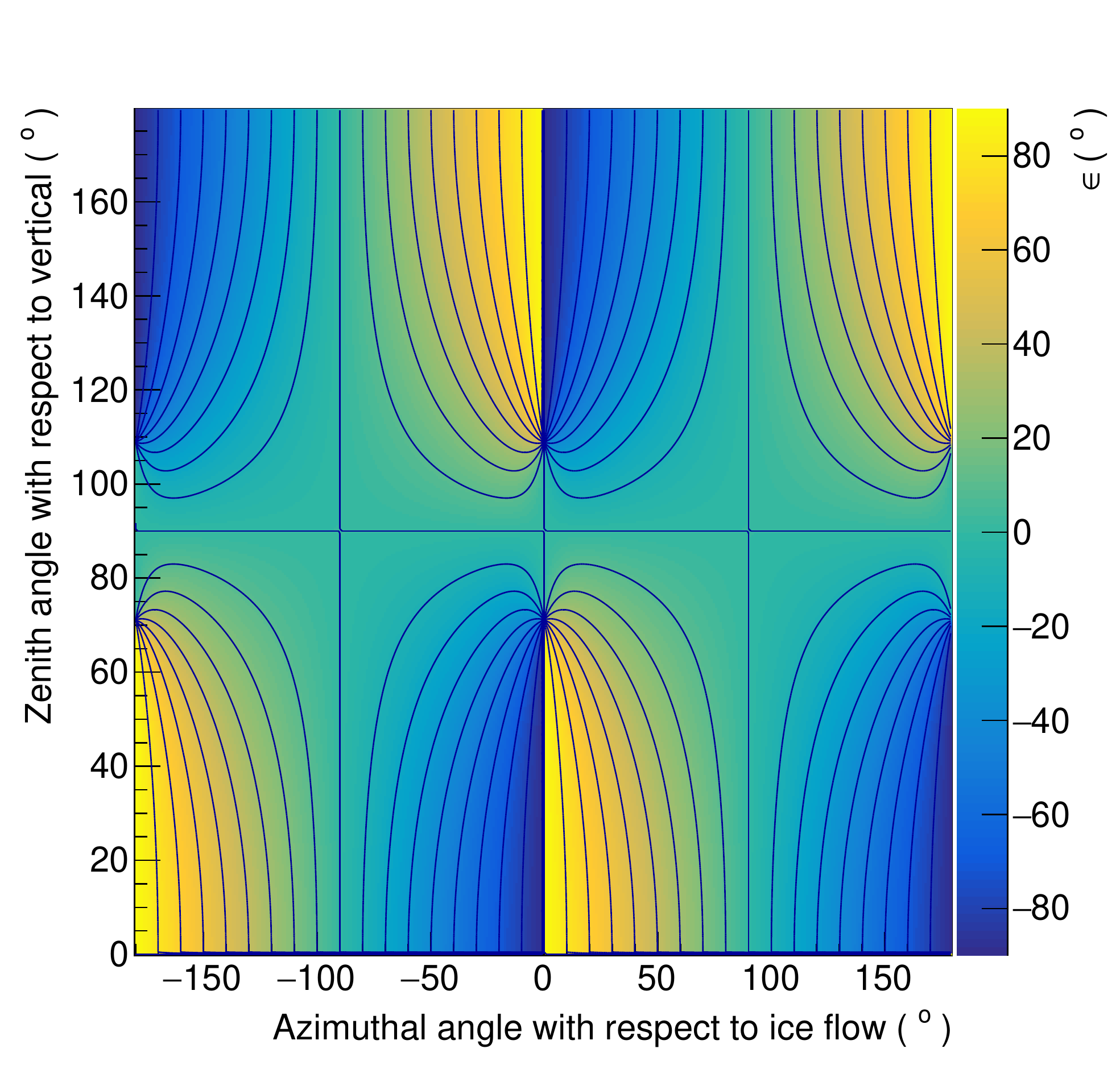}
    \caption{Angle $\epsilon_1$ as a function
    of direction of ${\bf k}$ at 200\,m depth
    (left) and 1700\,m depth (right). 
    Following Fig.~\ref{fig:n123}, at 200\,m depth,
$n_{\alpha}=1.77850$, $n_{\beta}=1.77900$, and
  $n_{\gamma}=1.78150$ are used for this figure, and at 1700\,m,
  $n_{\alpha}=1.77700$, $n_{\beta}=1.78125$, and
  $n_{\gamma}=1.78175$ are used.
  The vertical and horizontal lines correspond to directions that lie in the
  $\alpha-\beta$, $\beta-\gamma$, or the $\alpha-\gamma$ plane.
  The points where the contours meet correspond to the lines
  in the $\alpha$-$\gamma$ plane where the intersection ellipse
  becomes a circle and the directions of the eigenvectors are undefined.  These occur at zenith angles equal to the $V_z$ angle
  in Fig.~\ref{fig:V}.  At 200\,m depth, $V_z$ is about 20$^{\circ}$ 
  and at 1700\,m depth it is approximately 65$^{\circ}$, corresponding to the zenith angle of the 
  points where the contours converge in
  the left the right plots, respectively.}
    \label{fig:epsilon_vs_k}
\end{figure*}

Fig.~\ref{fig:epsilon_vs_k} shows the
epsilon angle $\epsilon_1$ as a function of the direction of the ${\bf k}$
expressed in zenith and azimuthal
angles, for 200\,m and 1700\,m depths.  The contours are lines of
equal $\epsilon$.  The vertical and
horizontal lines are directions where ${\bf k}$ is in a plane containing
the $\alpha-\beta$, $\beta-\gamma$,
or $\alpha-\gamma$ axes.  We can see
that in those planes the $\epsilon$ angles go to zero, but in the $\alpha$-$\gamma$ plane, the $\epsilon$
angles can change rapidly in directions
deviating from that plane. 

In the $\alpha$-$\gamma$ plane, there is a direction of ${\bf k}$ where the
directions of the eigenvectors become
indeterminant and the contours
converge.  This corresponds to the
direction in which the intersection
of the planar wavefront with the
indicatrix becomes a circle, and so
the axes of the intersection ellipse
that normally define the directions
of the eigenvectors are undefined.
This is the same as the $V_z$ angle in
Fig.~\ref{fig:V}.  For example,
at 200\,m depth, the contours
converge at $\theta\approx 20^{\circ}$
in the $\alpha$-$\gamma$ plane,
and at 1700\,m depth at $\theta \approx 70^{\circ}$, which are the values of
$V$ at those depths in Fig.~\ref{fig:V}.

\label{sec:epsilon_vs_k}

\section{Orientation of the
principal axes}

Many past
investigations related
to radio-frequency birefringence in polar ice sheets, including previous studies
interpreting SPICE pulser data in the context
of birefringence in
South Pole ice
mentioned in Sec.~\ref{sec:introduction},
have assumed that the indicatrix 
is oriented with
the $\gamma$-axis  vertical and another principal axis in the direction of
ice flow.  However,
measurements in other locations
have found that these assumptions
do not hold at the $\sim10^{\circ}$ level, 
which is likely also true at South Pole.

The ``tilt'' angle that the $\gamma$-axis makes with the vertical direction
can be obtained from either measurements
of ice cores or radar measurements.
Matsuoka {\it et al.}~\cite{Matsuoka2009} summarizes
results of core measurements where the tilt
angle is several degrees at many
sites in both Greenland and
Antarctica.  J. Li {\it et al.}~\cite{tc-12-2689-2018}
used multi-polarization radar
measurements at the NEEM site
to ascertain a tilt angle of
9.6$^{\circ}$ from the vertical axis.

Typically when ice cores
are extracted, the azimuthal
angle of the core is not preserved.
Radar measurements do
have the ability to extract
this orientation.
For example,
Jordan {\it et al.}~\cite{Jordan:2019bqu} found
the direction of the $\beta$-
axis at the
North Greenland Eemian Ice
Drilling (NEEM) ice core region in Greenland
to be as much as 25$^{\circ}$
away from its nominally expected
direction perpendicular to flow, and
that it changed by about
10$^{\circ}$ between sites separated
by a few km.
T.J.~Young {\it et al.}~\cite{tc-15-4117-2021} estimated
the ${\alpha}$-axis to be 14$^{\circ}$
from the direction of ice flow
at WAIS Divide in Antarctica.

As can be seen from Fig.~\ref{fig:epsilon_vs_k}, uncertainties on the orientation 
of principal axes
of order several degrees at South Pole would,
from some ${\bf k}$ directions, impact
the $\epsilon$ angles and thus the polarization
directions by of order 10$^{o}$, and this
would directly impact the ability to reconstruct the direction of a neutrino.
In Ref.~\cite{Jordan2019}, it was shown (see Fig.\,10) that tilt angles
of about 10$^{\circ}$ can lead to 20\% uncertainties on the
difference between $n_{\alpha}$ and $n_{\beta}$, which affects delay times and
thus distance and energy reconstruction.  Uncertainties in the orientation relative to ice flow would have a similar effect.


\bibliography{apssamp}

\end{document}